\documentclass[aps,prb,twocolumn,superscriptaddress,showpacs]{revtex4}

\usepackage{amssymb}
\usepackage{graphicx}
\usepackage{amsmath}
\usepackage{hyperref} 

\begin{document}

\title{Real-time simulation of finite frequency noise from a single electron emitter}

\author{T.~Jonckheere}
\email[Email address: ]{thibaut.jonckheere@cpt.univ-mrs.fr}
\affiliation{Centre de Physique Th\'eorique, CNRS UMR 6207, Aix-Marseille Universit\'e, Case 907, 13288 Marseille, France}
\author{T.~Stoll}
\affiliation{Centre de Physique Th\'eorique, CNRS UMR 6207, Aix-Marseille Universit\'e, Case 907, 13288 Marseille, France}
\author{J.~Rech}
\affiliation{Centre de Physique Th\'eorique, CNRS UMR 6207, Aix-Marseille Universit\'e, Case 907, 13288 Marseille, France}
\author{T.~Martin}
\affiliation{Centre de Physique Th\'eorique, CNRS UMR 6207, Aix-Marseille Universit\'e, Case 907, 13288 Marseille, France}

\date{\today}

\begin{abstract}
We study the real-time emission of single electrons from a quantum dot coupled to a one dimensional conductor, using exact diagonalization
on a discrete tight-binding chain. We show that from the calculation of the time-evolution of the one electron states, 
we have a simple access to all the relevant physical quantities in the system. In particular, we are able to compute
accurately the finite frequency current autocorrelation noise. The method which we use is general and versatile, allowing to study the impact of many different parameters like the dot transparency or level position. Our results can be directly compared with existing experiments, and
can also serve as a basis for future calculations including electronic interactions using the time dependent density-matrix renormalisation group 
and other techniques based on tight-binding models. 
\end{abstract}


\maketitle

\section{Introduction}
\label{sec:introduction}
The study of electronic transport in mesoscopic systems -- where small size, low temperature and careful fabrication of the sample
ensure that quantum coherence is preserved -- has shown tremendous progress in the past decades, providing a deep understanding of the behaviour of these systems.\cite{sohn_mesoscopic_1997,houches2005} Recent progress opened the way to the study of these systems on short time-scales, by looking at either the real-time dynamics
 or the high-frequency fluctuations.\cite{onac_using_2006,zakka-bajjani_experimental_2007,aguado_double_2000,zazunov_detection_2007}
 A recent experiment has demonstrated the feasibility of single electron emission, where 
a single electron is periodically emitted from a quantum dot into an edge state of a 2d electron gas in the quantum Hall effect \cite{feve_-demand_2007} (see also Ref. \onlinecite{leicht_generation_2011}).
Measurements of the current and finite frequency current correlations in this system confirmed that the system indeed behaves as a single-electron emitter.\cite{mahe_current_2010}

\begin{figure}[b]
\centerline{\includegraphics[width=8.cm]{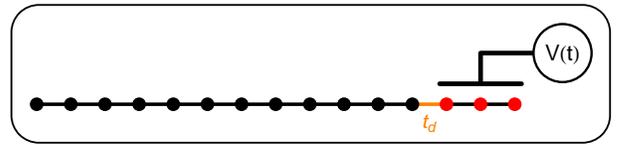}}  
\caption{{(color online) Sketch of the system: a tight-binding chain of length $L$, with a dot defined on one end (red sites, at the
right of the $t_d$ link), on which a time-dependent
potential $V(t)$ is applied.}}
\label{fig:chaingraph}
\end{figure}

In this work, we study this system using real-time numerical simulations. Modeling the edge state and the quantum dot as a tight-binding chain
without electronic interactions (see Fig.~\ref{fig:chaingraph}),
we compute numerically the time evolution of the system, from that of the one-electron states, when
a time-dependent gate voltage is applied to the dot. We are able to study all the aspects of single-electron emission, by calculating  the
average of all the relevant physical operators. In particular, we show that these real-time calculations allow us to
compute accurately the finite frequency autocorrelation noise of the emitted current, which has been used as the definite experimental proof
of single-electron emission.\cite{mahe_current_2010} 
In the optimal emission regime, our method provides very good agreement with the analytical results for the finite frequency noise obtained from a semi-classical model\cite{albert_accuracy_2010}, which proves the power of real-time simulations in this context. More importantly, our approach can be used to study several different regimes where no analytical results are available.

The goal of this work is twofold. First, as the single-electron emission experiments are currently performed in a regime where electronic interactions in the one-dimensional channel are small and can be neglected, the results obtained within our real-time simulation for non-interacting electrons are relevant as such, and can be directly compared with experimental data. We stress however that fermionic correlations are accounted for exactly with our approach.
The method is quite versatile, and allows to easily study the effect of various parameters, like the transparencies, the temperature, the exact shape of the driving voltage, etc. Moreover, the real-time simulation gives access to appealing
visualisation of the electronic transport in these systems, with easy access to the real-time density in real and energy space, as well as the shape of emitted wavepackets. As a means to get results for the non-interacting system, our method thus appears as complementary to calculations based on the Floquet scattering 
theory\cite{moskalets_floquet_2002, moskalets_time-resolved_2007, moskalets_quantized_2008,grenier_single-electron_2011,parmentier_current_2011},
 with equivalent results but different strengths and weaknesses.

Second, this work is a necessary first step towards the use of more
involved real-time simulations, such as
time-dependent density matrix renormalization group (td-DMRG) and related techniques,\cite{white_density_1992,
cazalilla_time-dependent_2002,white_real-time_2004,daley_time-dependent_2004,dias_da_silva_transport_2008,heidrich-meisner_real-time_2009,
branschaedel_conductance_2010,ulrich_density-matrix_2011}
where electronic interactions can be taken into account, in such a time-dependent situation.

 The understanding of the role played by the different timescales, finite size effects and  discretization, which our approach provides for non-interaction electrons, constitutes a prerequisite for the application of these more advanced and complex methods. 
 In this regard, the present work is made in the same spirit
 as Ref.~\onlinecite{branschaedel_numerical_2010} (where shot-noise was computed from real-time dynamics) but actually goes beyond in considering the finite-frequency noise in a time-dependent setup.

This article is organized as follows. In Sec.~\ref{sec:model}, we detail the model which we have used, and we derive expressions for the averages
of physical quantities in terms of the numerical solution for the time evolution of the one-electron states. 
In sec.~\ref{sec:results}, we present our results, first considering the static properties of the dot (mean charge), and then extending these considerations to the time-dependent properties (emitted charge density along the chain, 
emitted current, and autocorrelation noise). Sec.~\ref{sec:conclusion} is devoted to discussions and conclusions.

\section{Model}
\label{sec:model}

\subsection{General formalism}
\label{sec:formalism}
\begin{figure}
\centerline{\includegraphics[width=6.cm]{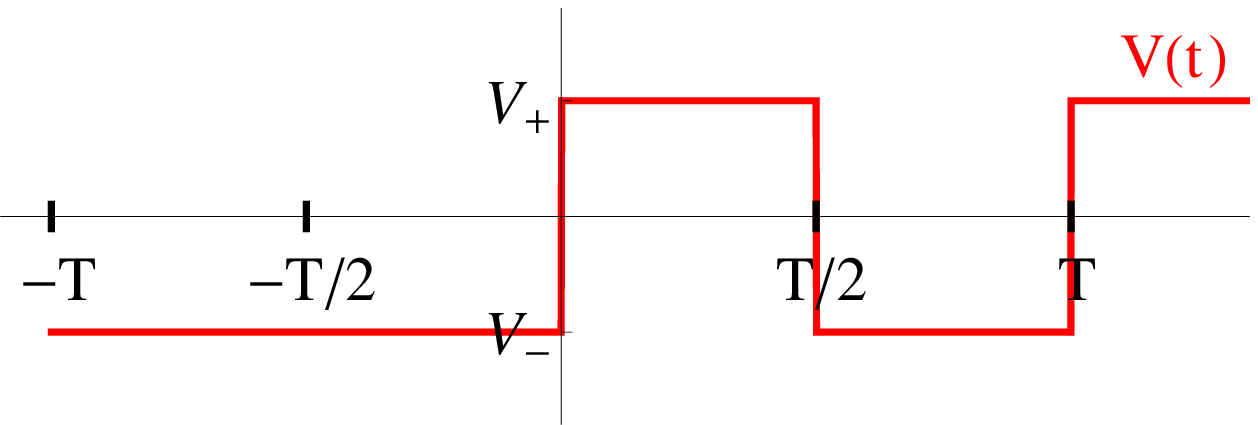} } 
\centerline{\includegraphics[width=6.5cm]{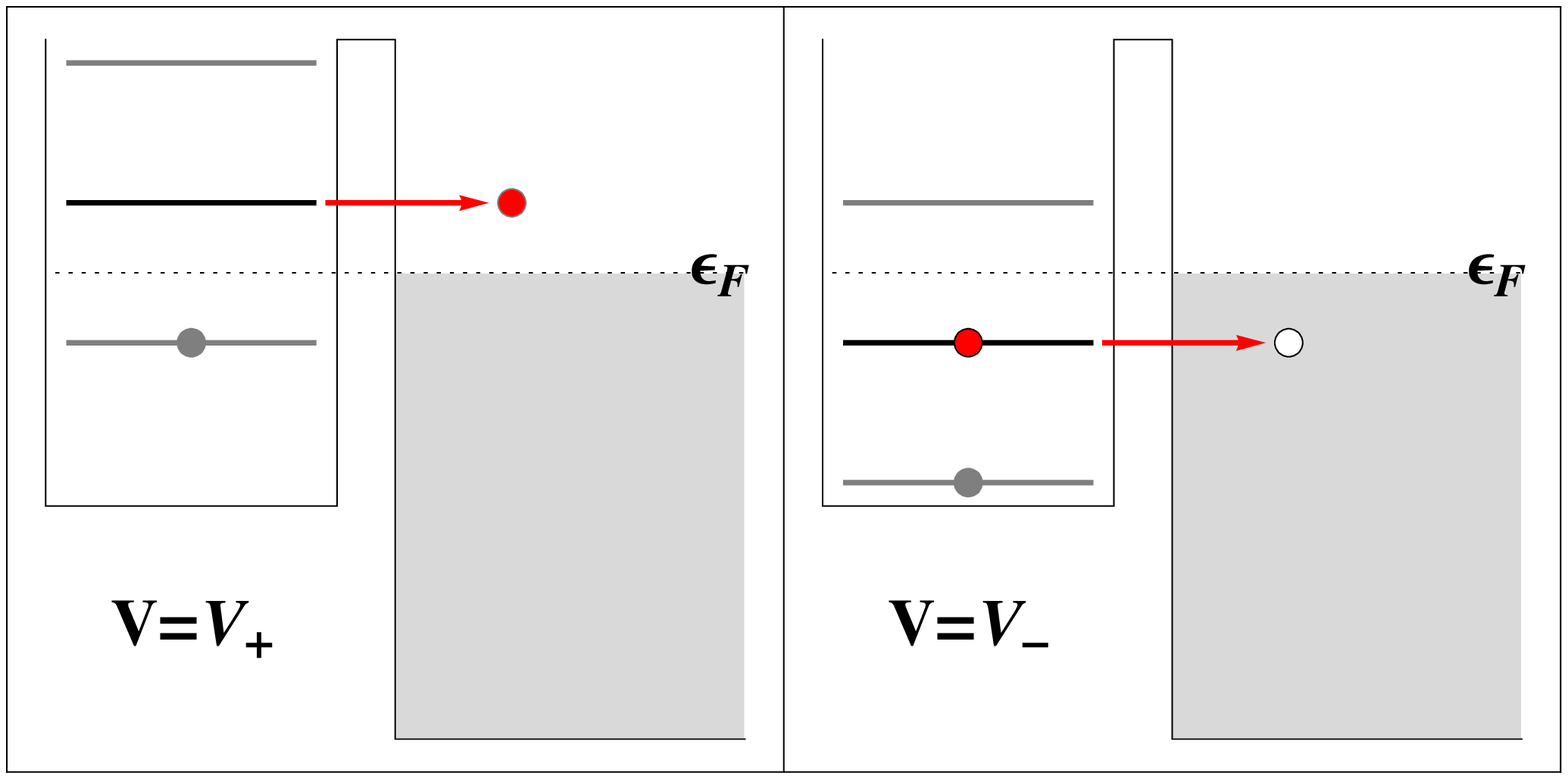} } 
\caption{(color online) Top: Profile of the potential $V(t)$ used in our calculations: $V(t)$ is constant and equals $V_{-}$ for $t\leq 0$
 (system in equilibrium), then shows perfect periodic steps between $V_{+}$ and $V_{-}$ for $t>0$.  
 Bottom: illustration of the electron emission process in the optimal regime, 
 where $V_{+} - V_{-} = \Delta$ (the level spacing of the dot), and the values
 of $V_{+}$ and $V_{-}$ are chosen such that the Fermi energy $\epsilon_F$ of the chain is always in the middle of two energy levels. The
 left drawing shows the electron emission, when the driving potential brings an occupied level (black) above the Fermi energy; the right
 drawing shows the hole emission, when the empty level (black) is put below the Fermi energy by the driving potential.
 }
\label{fig:Vt}
\end{figure}

The system is modeled as a one dimensional (1d) tight-binding chain of length $L$, with the Hamiltonian:
\begin{multline}
\label{eq:hamiltonian}
H = -\sum_{j=1, \neq j_d}^L \left(\hat{c}_j^{\dagger} \hat{c}_{j+1} + \mbox{h.c.} \right)
- t_d \left(\hat{c}_{j_d}^{\dagger} \hat{c}_{j_d+1} + \mbox{h.c.} \right) \\
+(\epsilon_d + V(t)) \sum_{j>j_d} \hat{c}_j^{\dagger} \hat{c}_j
\end{multline}
where $\hat{c}_j$ and $\hat{c}_j^{\dagger}$ are fermionic annihilation/creation operators at site $j$. The dot is composed of the sites $j>j_d$, with a tunneling amplitude $t_d$ to/from the dot, an on-site energy $\epsilon_d$, and an applied time-dependent potential $V(t)$. 
To perform periodic single-electron emission, 
$V(t)$ is a periodic function of time (with period $T$), which in the optimal situation brings alternatively the highest occupied
dot level above the Fermi energy in the leads (electron emission), and the lowest unoccupied state below the Fermi energy 
(hole emission), see Fig.~\ref{fig:Vt}.
 In order to compute the evolution of the system, we consider it to be in equilibrium at $t \leq 0$ with $V(t)=\mbox{const}$, while the periodic potential is applied for $t>0$.  

For $t \leq 0$, the system is in equilibrium, 
and we can compute the $L$ eigenstates of the Hamiltonian $H$, noted $\phi_i$ ($i=1 \dots L$) 
with increasing energies $E_1 < E_2 < \dots < E_L$.
The initial $N$-particle state, at zero temperature, is simply obtained by filling the $N$ lowest energy states. For non-zero
temperature, the occupation of the states is given by the Fermi distribution.
 In the following, we will work with the chain at half-filling, with $N= L/2$.
  For $t>0$, the evolution of each one-electron states $\phi_i(t)$ can be obtained
by solving the time dependent Schr\"{o}dinger equation: 
\begin{equation}
i \hbar \frac{\partial \phi_i(t)}{\partial t} = H(t) \; \phi_i(t)~.
\label{eq:Schrod1el}
\end{equation}
This can be achieved by numerical integration of the differential equation for an arbitrary potential $V(t)$.
  
As there are no electronic interaction in the system, each occupied electronic state evolves independently form the
other states, and the $N$-particle state at time $t>0$ is simply obtained by filling the same 
1-electron state as for the initial state, but using now the time-dependent $\phi_i(t)$.
 Introducing $\hat{\phi}_i^{\dagger}(t)$ as the fermionic operator creating an electron in the single-electron state $i$ at time $t$,
the $N$-particle wavefunction of the chain at time $t$ is:
\begin{align}
\left| \Psi(t) \right \rangle &= \hat{\phi}_1^{\dagger}(t) \; \hat{\phi}_2^{\dagger}(t)\; \dots\;
                                   \hat{\phi}_{L/2}^{\dagger}(t) \;\; | 0 \rangle \nonumber \\
                              &\equiv \left| 1,1,\dots,1,0,0,\dots,0 \right \rangle_t~,
\label{eq:Psit}
\end{align}  
where $|0\rangle$ denotes the empty band.

We want to stress that this method does not rely on any adiabatic approximation for the evolution due to $V(t)$,
 and it can be applied with an arbitray potential $V(t)$. In an nutshell,
  what we are doing is simply to take the system in a given initial state ($t<0$), 
  and to compute its time-evolution using the time-dependent Schr{\"o}dinger equation. This can be done
 in principle for any system (e.g. even with electronic interactions),
  but would be impractical at the level of the $N$-electron state because of the huge
 size of the Hilbert space. But since we are considering here
 non-interacting electrons, this can be done for each electron independenlty (Eq.(\ref{eq:Schrod1el})), thus relatively easily.

For the sake of simplicity, we consider here the special case of a piecewise constant potential $V(t)$, which allows to reduce the numerical integration of the differential equation to simple matrix products (see below).
Note however that the numerical cost for solving numerically the differential equation with a more complex $V(t)$ is only slightly higher, so that in practice our method can
 be applied without difficulty to any reasonable form of $V(t)$.
We thus focus on a potential $V(t)$ which at $t>0$ consists of perfect periodic steps (see fig.~\ref{fig:Vt}) and takes the form:
\begin{equation}
V(t) =  
 \begin{cases}
 & V_{-} \quad \mbox{if} \quad t\leq 0 \\
 & V_{+} \quad \mbox{if} \quad 0 + n T \leq t<\frac{T}{2} + n T  \\ 
    &  V_{-} \quad \mbox{if} \quad \frac{T}{2} + nT < t \leq T + n T 
  \end{cases}  
\end{equation}
where $n$ is a positive integer. In this case, knowing the eigenvalues and eigenstates of the system for $V=V_{-}$ and $V=V_{+}$ is enough to get the time evolution of any one-particle state, by expressing it alternatively in the basis of these
eigenstates. During a time interval where $V(t)$ is constant, the time evolution is simply given by phase factors coming from the eigenenergies. Then the sudden switch of $V(t)$ from $V_{+}$ to $V_{-}$ (or vice-versa) is accounted for by a basis transformation
from the $V_{+}$ eigenstates to the $V_{-}$ eigenstates (or vice-versa), which ultimately amounts to a simple matrix product. By combining
the trivial time-evolution for a constant $V(t)$ with the change of basis when $V(t)$ switches from one value to another, we easily get access to the one-particle state evolution for any time $t$.

\subsection{Calculation of operators average}

The fermionic operator for the one-particle state $\hat{\phi}_i(t)$
can be written in full generality in the basis of the position states along the chain:
\begin{equation}
\label{eq:DefAlpha}
\hat{\phi}_i(t) = \sum_{j=1}^{L} \alpha_{i,j}(t) \hat{c}_j ,
\end{equation}
where $\alpha_{i,j}(t)$ is a time-dependent unitary matrix. This matrix is the natural object coming out of the numerical
evolution of the system, as it expresses the one-particle states at time $t$ in terms of the position states.
In particular, the lines of the matrix $\alpha_{i,j}(t)$ at $t=0$ correspond to the eigenstates of the Hamiltonian (\ref{eq:hamiltonian}), which
are naturally obtained in the position state basis, as the Hamiltonian itself.    

For convenience, we compute all the operator averages in the Heisenberg picture. The time-dependent position operators along 
the chain can then be written as:
\begin{equation}
\hat{c}_k(t) = \sum_{m=1}^L \alpha_{m,k}^*(t) \hat{\phi}_m(0)
\end{equation}
where the one-electron states operators are taken at $t=0$.

\subsubsection{Average current}

The current operator on link $k$ (between sites $k$ and $k+1$) at time $t$ is:
\begin{equation}
\hat{I}_k(t) = -i \left( \hat{c}_k^{\dagger}(t) \hat{c}_{k+1}(t) - \mbox{h.c.} \right) 
\end{equation}
and its average is:
\begin{multline}
\left \langle I_k(t) \right \rangle = -i \Big( \sum_{m,n} \alpha_{m,k}(t) \alpha_{n,k+1}^{*}(t)  
     \\ \left \langle \Psi(0) \middle| \hat{\phi}_m^{\dagger}(0) \hat{\phi}_n(0) \middle|  \Psi(0) \right\rangle - \mbox{h.c.} \Big)
\end{multline}
with $| \Psi(0) \rangle\equiv | \Psi\rangle$ the half filled Fermi sea at time $t=0$.
Using Eq.~\eqref{eq:Psit}, at zero temperature,
 the average of the $\hat{\phi}$ operators product is simply $\delta_{m,n}$ for the occupied states ($m\leq L/2$), and $0$ for the empty ones ($m>L/2$), leading to:
 \begin{equation}
 \label{eq:currentT0}
 \left \langle I_k(t) \right \rangle = 2 \sum_{n\leq L/2} \mbox{Im} \left( \alpha_{n,k}(t) \alpha_{n,k+1}^*(t) \right) .
 \end{equation}
At non-zero temperature, one has $\langle \hat{\phi}_n^{\dagger} \hat{\phi}_n \rangle = f(E_n)$, where $f(E)$ is the
 Fermi function at energy $E$, and the average current becomes:
  \begin{equation}
 \label{eq:CurAv} 
 \left \langle I_k(t) \right \rangle = 2 \sum_{n=1}^{L} f\left(E_n \right) \mbox{Im} \left( \alpha_{n,k}(t) \alpha_{n,k+1}^*(t) \right)  \;.
 \end{equation}   
 
\subsubsection{Current correlations} 
The correlation between the current at link $k$ and time $t_1$ with the one at link $l$ and time $t_2$ is given by:
\begin{multline}
S_{k,l} (t_1,t_2)=\left \langle \Psi \middle| I_k(t_1) I_l(t_2) \middle| \Psi \right \rangle 
 \\ -\left \langle \Psi \middle| I_k(t_1) \middle| \Psi \right \rangle
    \left \langle \Psi \middle| I_l(t_2) \middle| \Psi \right \rangle .
 \end{multline}  
 Proceeding along the same lines as for the current average, and using Wick's theorem for averages involving four $\hat{\phi}$ operators,
 we get at finite temperature:
  \begin{multline}
  \label{eq:NoiseAv}
 S_{k,l} (t_1,t_2) =
  (-1) \sum_{m=1}^{L} \sum_{n=1}^L  f(E_m) \left( 1- f(E_n) \right)
   \\ \times \left( \alpha_{m,k}(t_1) \alpha^*_{n,k+1}(t_1) - \alpha_{m,k+1}(t_1) \alpha^*_{n,k}(t_1)\right)  \\
  \quad \times \left( \alpha_{n,l}(t_2) \alpha^*_{m,l+1}(t_2) - \alpha_{n,l+1}(t_2) \alpha^*_{m,l}(t_2)\right) .
  \end{multline} 

\subsubsection{Mean charge on the dot}
The operator $\hat{Q}(t)$ describing the charge on the dot is:
\begin{equation}
\hat{Q}(t) = \sum_{j>j_d} \hat{c}^{\dagger}_j(t) \hat{c}_j(t) ,
\end{equation}
leading to the average charge:
\begin{align}
\label{eq:ChargeAv}
\langle Q(t) \rangle & = \sum_{j>j_d} \sum_{m,n} \alpha_{m,j}(t) \alpha_{n,j}^{*}(t)  \nonumber \\
&\quad \quad \quad \times \left \langle \Psi(0) \middle| \hat{\phi}_m^{\dagger}(0) \hat{\phi}_n(0) \middle| | \Psi(0) \right\rangle  \nonumber
\\ &= \sum_{j>j_d} \sum_{n=1}^{L} f(E_n)  \left|\alpha_{n,j}(t)\right|^2  .
\end{align}

The results from Eqs.~(\ref{eq:CurAv}), (\ref{eq:NoiseAv}) and (\ref{eq:ChargeAv}) show that all the averages can easily be computed from the time-dependent matrix $\alpha_{i,j}(t)$, which is directly obtained from the numerical computation of the time evolution. 
 
\section{Results}
\label{sec:results}

\subsection{Parameters and operating regime}

We present below the results which we have obtained to characterize the emission of single electrons in the one-dimensional chain. 
One must be aware that our approach uses a non-chiral system, where excitations can propagate to the left and to the right, in order to simulate a chiral system (an edge state of the quantum Hall effect). In our setup, the left-going excitations correspond to the ones coming out of the dot in the real system, while the right-going excitations are incoming onto the dot, and thus need to be avoided. Multiple reflections inside the simulated dot are equivalent in the real system to performing multiple round trips inside the dot.

In the remainder of the text, we work with a dot composed of $5$ sites, located at the end of a much longer chain ($\sim$ a few hundred sites, typically 500). This size of the dot is the result of a compromise. On the one hand, the dot has to be as small as possible compared to the chain, in order to minimize finite-size effects. Indeed the level spacing in the chain must be negligible compared to the level spacing in the dot, to appear
as a continuum. On the other hand, since we apply a time-dependent voltage of the order of the dot level spacing, the latter needs to be small compared to the bandwidth of the tight-binding chain, in order to minimize the effects of the non-linear dispersion relation of the chain. As the level spacing is inversely proportional to the size of the
dot, one thus needs to consider a large enough dot to meet this requirement. In practice, we found that using a dot of 5 sites was a good compromise, in particular because we were  able to perform the current measurement near the dot, thus reducing the spreading of wavepackets during propagation, which arises from the non-linear dispersion.

We consider the chain to be at half-filling, and work with physical dimensions corresponding to $\hbar= e =1$. The tunneling amplitude along the chain (first term in the Hamiltonian (\ref{eq:hamiltonian})) has been taken as the unit of energy, and as a consequence the Fermi velocity at half-filling is equal to 2 sites per unit of time.
  This means that an excitation takes a time $L$ to reflect at the boundary and come back to its starting point along a chain of $L$ sites. In particular, the current measured at a given link due to the passage of an excitation will get completely spoiled after a time $L$ by the reflection of that same excitation at the end of the chain. One can thus expect that the numerical quantities which we compute will be reliable up to a time $\sim L$ only.

Unless specifically stated, all the results presented below were obtained in the optimal regime for electron emission. In this regime,
the amplitude of the driving potential, $V_{+}-V_{-}$, is taken to be equal to the dot level spacing $\Delta$.
 When the driving potential switches to the value $V_{+}$,  one energy level of the dot is put at an energy $\epsilon_F+\Delta/2$ 
 (where $\epsilon_F$ is the Fermi energy of the chain), and an electron is emitted at this very energy 
 (see the bottom left panel of Fig.~\ref{fig:Vt}). When $V(t)$ switches back to $V_{-}$, this same level is brought down to the
  energy $\epsilon_F-\Delta/2$, and a hole is emitted, i.e. an electron tunnels back to the dot level 
  (see the bottom right panel of Fig.~\ref{fig:Vt}). This parameter regime is the best suited to emit a true single electron and a single
  hole during each period, as has been shown experimentally\cite{feve_-demand_2007}, and it is thus the most relevant one for our study.   
  Note however that there is no difficulty in exploring other operating regimes with our method (for example, a dot level in resonance with $\epsilon_F$).

\subsection{Static properties of the dot}

\begin{figure}
\centerline{\includegraphics[width=7.cm]{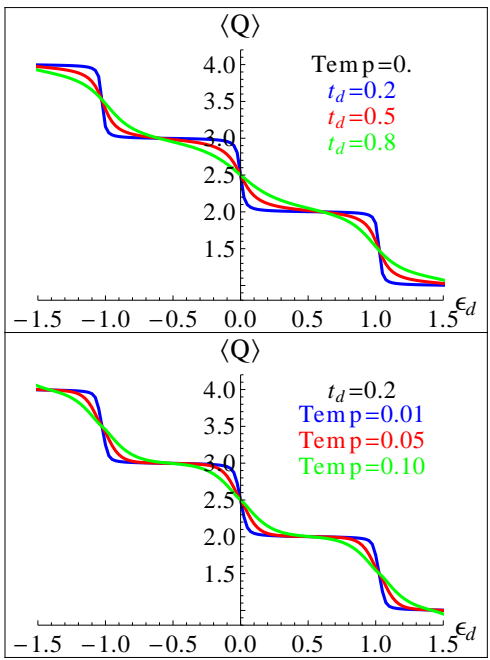}} 
\caption{(color online) Mean charge on the dot as a function of the on-site energy of the dot $\epsilon_d$, for several tunneling
amplitudes to the dot $t_d$ at zero temperature (top), and for several temperatures at $t_d=0.2$ (bottom). For small $t_d$ and 
temperature, the mean charge tends to be quantized at integer values, with sharp steps between the plateaus. When $t_d$
or the temperature are increased, the steps are smoothed due to the increased fluctuations between the dot and the lead. }
\label{fig:Qstatic}
\end{figure}

Before studying the properties of the electron emission by the dot, it is necessary to characterize the properties of the dot
itself without a time-dependent potential $V(t)$ applied. Fig.~\ref{fig:Qstatic} shows the mean charge on the dot as a function of the
on-site energy $\epsilon_d$, for several values of the tunneling amplitudes to the dot $t_d$ at zero temperature (top plot),
and for increasing temperature with $t_d=0.2$ (bottom plot). For a small value of $t_d$ at zero temperature, we observe that the mean 
charge is quantized, with well-defined plateaus at integer value. This simply reflects that when the dot is weakly coupled to the chain, the
number of electrons on the dot is an integer: it can vary between 5 and 0, leading to five energy levels. Indeed, the density of states
(not shown) which corresponds to $\partial \langle Q \rangle / \partial \epsilon_F=-\partial \langle Q \rangle / \partial \epsilon_d$ has five narrow peaks at the position of the steps
in $\langle Q \rangle$, corresponding to the five energy levels of the dot. As $t_d$ is increased,
the steps between the plateaus are smoothed due to the increased fluctuations between the dot and the chain,
leading to broadened energy levels. Increasing the temperature for fixed $t_d$ (bottom plot) has a similar effect: 
it broadens the energy levels, leading to smoothed steps. 

For the parameters which we have chosen, the level spacing of the dot $\Delta$ is of the order of $1$. 
Yet, because of finite size effects, the level spacing $\Delta$ is not constant and depends on the levels considered.
 We thus focus on the electron/hole emission from the central level
of the dot ($\epsilon_d=0$), for which the optimal emission regime is obtained by taking $V_{+}=-V_{-}\simeq 0.6$. 

\subsection{Time-dependent charge on the dot}
\label{sec:Qt}
\begin{figure}
 \centerline{\includegraphics[width=8.5cm]{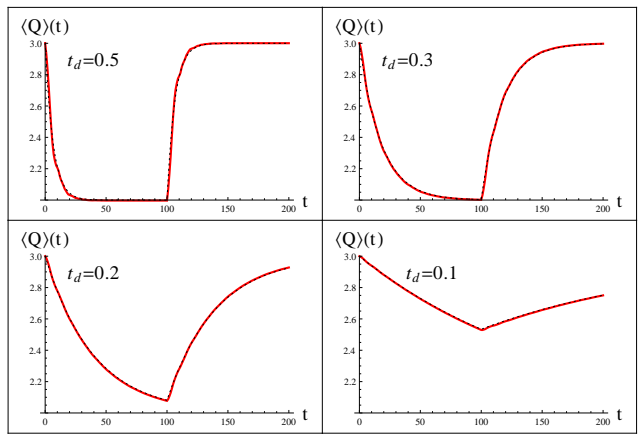}} 
 \caption{(color online) Mean charge $\langle Q \rangle$ on the dot as a function of time when $V(t)$
  is applied for different tunneling to the dot $t_d$
 at zero temperature. This data is obtained for a chain of length $L=500$, with a period $T=200$.}
 \label{fig:Qdynamic}
 \end{figure}

We consider now the time-dependent charge on the dot when $V(t)$ is applied, in the optimal emission regime 
($\epsilon_d=0$, $V_{+}=-V_{-}=0.6$).  Fig.\ref{fig:Qdynamic} shows the mean charge for this choice of
$V_{+}, V_{-}$ as a function of time ($0< t< T$), for several values of the tunneling amplitude to the dot $t_d$, at zero temperature. 
Several observations can be made from these plots.

 First, when the potential $V$ is switched from $V_{-}$ to $V_{+}$ (at $t=0$), one electron leaves the dot, and
the charge, up to a very good approximation, decreases exponentially: 
\begin{equation}
\langle Q \rangle (t) \simeq 2 + \mbox{exp}(-t/\tau)~,
\end{equation}
with a characteristic time $\tau$ which increases as $t_d$ decreases.
Calculations in the continuous limit for a perfect dot predict:\cite{feve_-demand_2007}
\begin{equation}
\tau \simeq \frac{2\pi}{\Delta} \frac{1}{D}, \quad \quad D \ll 1,
\end{equation}
where $D$ is the transparency of the dot. For the tight-binding chain which we consider,
 the relation between the tunneling amplitude $t_d$ and the
transparency is, in the linear regime of voltage:\cite{bardeen_tunnelling_1961,ferrer_contact_1988,PhDCuevas}
\begin{equation}
D(t_d)\simeq \frac{4 t_d^2}{\left(1+ t_d^2\right)^2} .
\end{equation}
Using this expression of $D(t_d)$ and performing a fit of the data gives the function $\tau \simeq 6.1/ D(t_d)$, with
a reasonable agreement with the value of the level spacing in our system  ($6.1 \simeq 2\pi/\Delta$ for $\Delta \simeq 1$).

When the potential $V(t)$ switches back to $V_{-}$ (for $t \geq T/2$), the charge starts to increase (as an electron comes back inside the dot, or equivalently a hole comes out of the dot), and the behaviour is symmetrically reversed, 
with a similar exponential growth back towards the initial charge.
 
 Second, the qualitative behaviour of the system changes as the ratio of the decay time $\tau$ to the half-period $T/2$ changes.
 When $\tau \ll T/2$, the electron is emitted/absorbed from the dot much faster than the half-period, and the emitted/absorbed charge is close to 1. 
 This is the case on the two upper plots of Fig.\ref{fig:Qdynamic}, where it is clear that the charge $\langle Q \rangle$ goes all the way from 3 to 2
 on the first half-period, then back from 2 to 3 on the second half-period. The electron is thus emitted or absorbed with a probability close
 to 1 at each half-period.
 On the other hand, when $\tau > T/2$, the electron emission/absorption is not complete during one half-period, which means that the probability
 that an electron is emitted or absorbed on each half-period is smaller than one. This is shown on the bottom plots of Fig.\ref{fig:Qdynamic}.

\begin{figure}
 \centerline{\includegraphics[width=8.5cm]{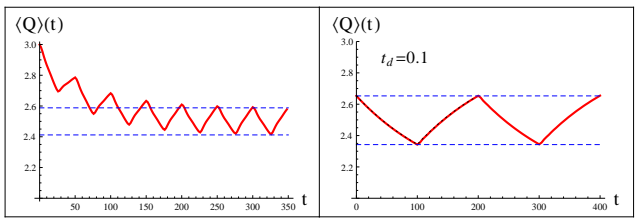}} 
 \caption{(color online) Mean charge $\langle Q \rangle$ on the dot as a function of time when $V(t)$ is applied,
 in the regime where $\tau \gg T/2$ (small tunneling amplitude to the dot).
 The left plot, obtained with a period $T=50$ on a chain of length $L=1000$ shows that starting at $t=0$ with $\langle Q \rangle = 3$ does not correspond to the long time limit: after several periods one reaches a long-time regime where $\langle Q \rangle$ oscillates
 between the values $2+Q_{+}$ and $2+Q_{-}$. The right plot, obtained with a period $T=200$ on a chain of length $L=500$,
  shows that it is possible to access the long-time limit from $t=0$ by switching on $V(t)$ at
 an earlier time $t_{in}<0$ to get the correct charge $\langle Q \rangle (t=0) = 2+Q_{+}$ (see text for details). }
 \label{fig:Qrelax}
 \end{figure}

 Note that, in this regime ($\tau > T/2$), the variation of the charge during one half-period is smaller than 1. One has
 then to be careful with the choice of initial conditions. Indeed, we are trying to model a periodic system using a finite one
 starting from an equilibrium/ground state configuration. 
 Depending on the initial parameters, the system will take one or several periods to relax towards the permanent
 periodic regime (which is independent of the initial conditions). This is illustrated on the left plot of Fig.\ref{fig:Qrelax},
  which shows $\langle Q \rangle(t)$, with $\tau \gg T/2$, for several periods (we chose $T=50$ for a chain of length $L=1000$,
  to be able to compute the evolution on many periods without suffering from reflections at the boundary of the chain). It is clear that the extrema of the oscillations of $\langle Q \rangle$ tend to relax towards long term values
  $2 + Q_{+}$ and $2 + Q_{-}$. As it is numerically very inefficient to compute the evolution over many periods (it requires very long chains to avoid boundary effects), 
  we have to modify the initial conditions in order to reach the periodic regime as soon as the time-dependent potential is applied.
  Knowing the value of the escape time $\tau$ from the dot for our choice of parameters,
  and introducing $s=\exp(-\frac{T}{2 \tau})$, the values of $Q_+$ and $Q_{-}$ can be predicted analytically,
   by solving the equations expressing that the charge at the end of one period (at $t=n T +T - 0$) has to be the same that the charge
   at the beginning of the period (at $t=n T + 0$):
 \begin{equation}
  Q_{-} = s \; Q_{+} \quad , \quad
  Q_{+} = s \; Q_{-} + \left(1- s \right)
 \end{equation}  
giving $Q_{+}= 1/(1+s)$, $Q_{-}= s/(1+s)$. In practice, in order to have the correct value of the charge ($2 + Q_{+}$)
 at $t=0$, we switch the potential $V(t)$ from $V_{-}$ to $V_{+}$ at a time $t_{in}<0$, such
that the value of $\langle Q \rangle$ at time $t=0$ is precisely $2+Q_{+}$. We are thus letting a fraction of electron escape
from the dot for $t<0$, in order for the charge to take the right value at $t=0$. The result of this procedure is shown on
 the right plot of Fig.\ref{fig:Qrelax}. One can see that the charge has indeed the predicted periodic behaviour, with increase/decrease
 between the values $2+Q_{+}$ and $2+Q_{-}$ shown as dashed lines. 
  
\subsection{Visualization of the time-dependent density}

\begin{figure*}
\includegraphics[width=8.cm]{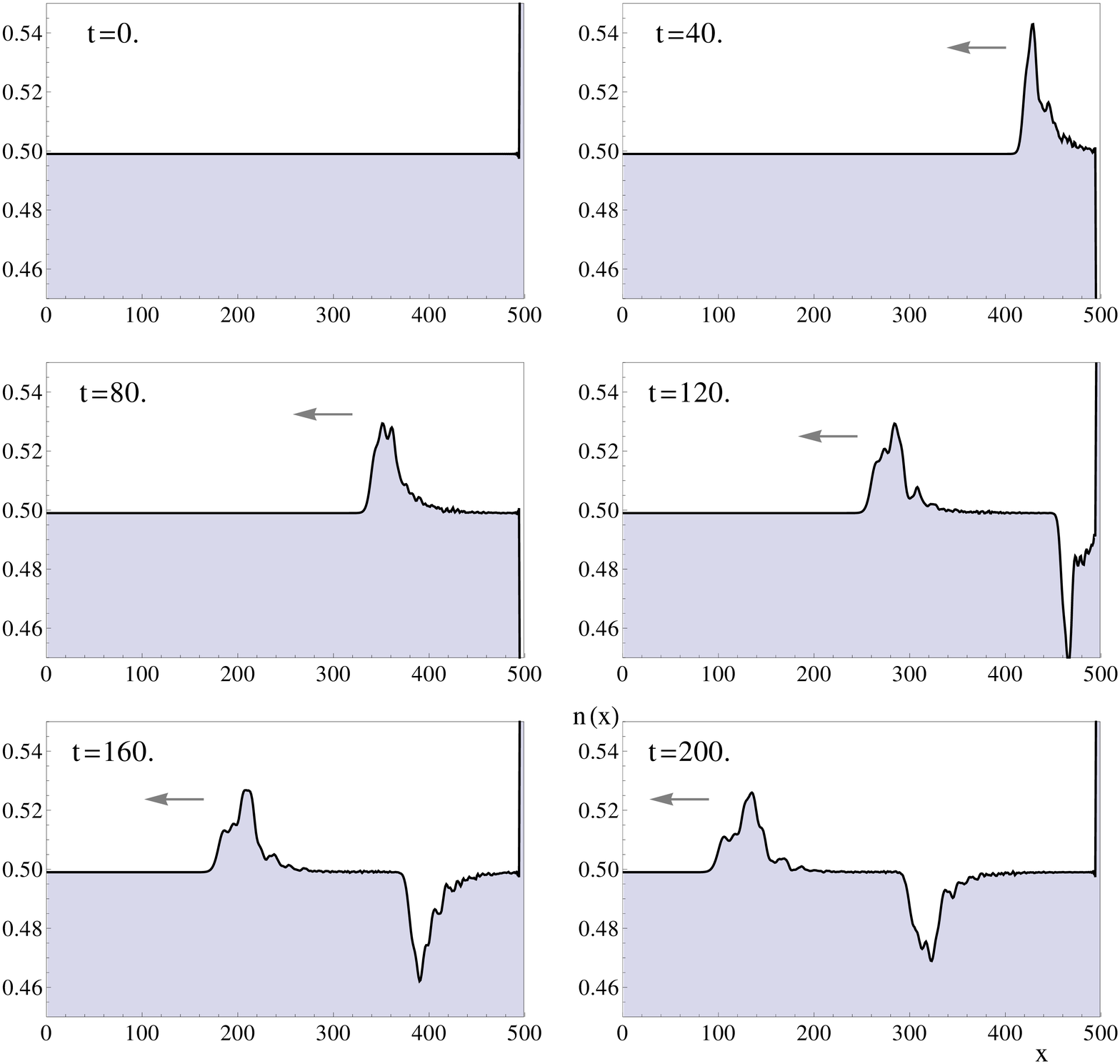} 
\hspace{1.cm}
\includegraphics[width=8.cm]{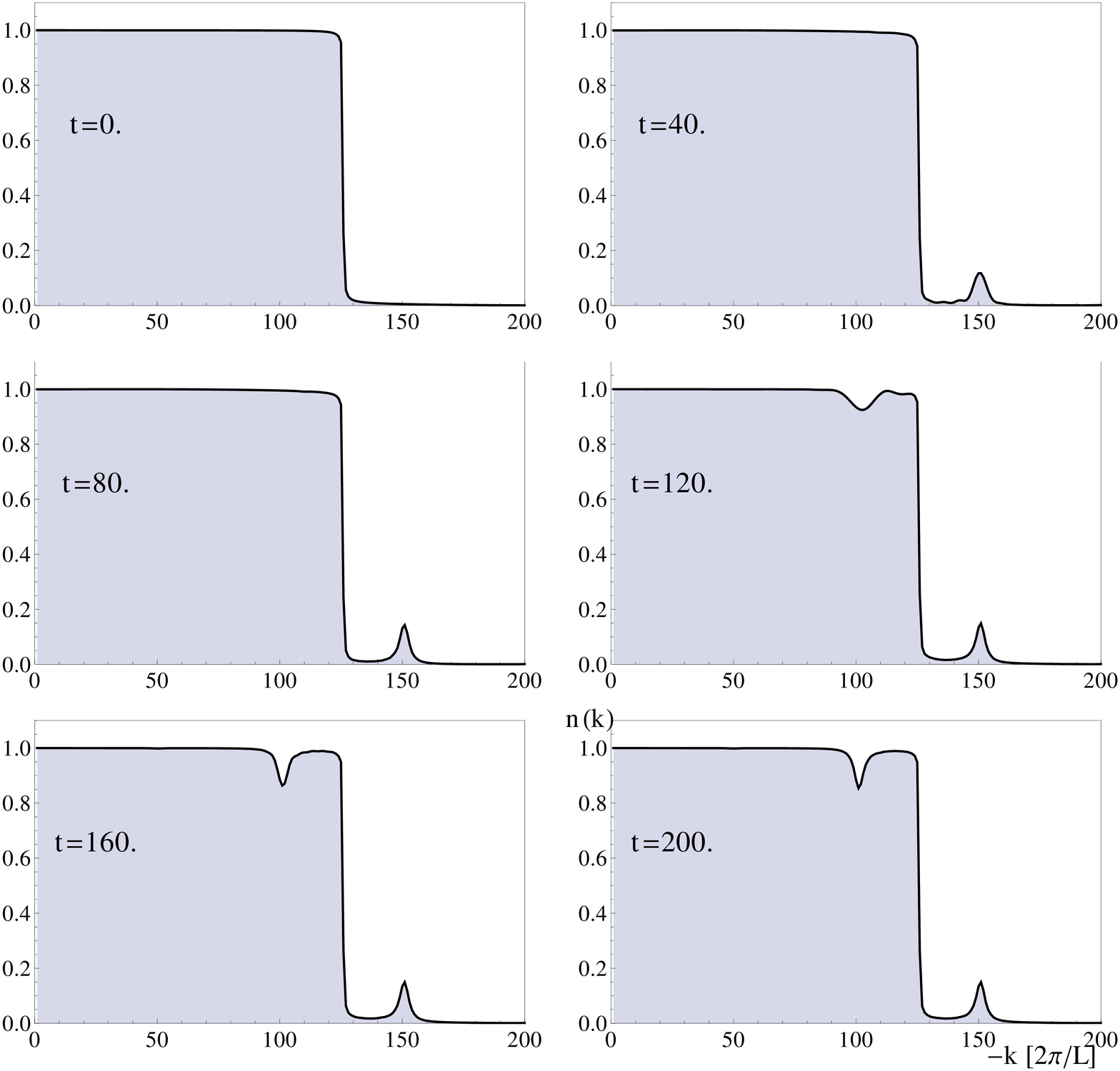} 
\caption{(color online) (Left) Density along the chain (mean occupation of each site) at different times 
for a chain of length $L=500$, with a dot tunneling amplitude $t_d=0.4$, at zero temperature, over one period of length $T=200$.
 One clearly sees the emission of the electron at $t=0$ with an exponential profile,
and of a hole at $t=T/2=100$. Because of the non-linear spectrum of the chain, wavepackets broaden during propagation.
(Right) Same evolution shown for the occupation in $k$ space, for negative $k$ (left-going excitation). 
The emission of the electron shows up as a well defined peak above the occupied states.} 
\label{fig:Density}
\end{figure*}
The numerical results allow to obtain a visual representation of the emitted wavepackets propagating along the chain.
The left part of Fig.~\ref{fig:Density} shows an example of results, obtained for a chain of length $L=500$, and an intermediate value
for the dot tunneling amplitude, $t_d=0.4$, at zero temperature. The chain is initially in equilibrium at half-filling (occupation =0.5
on all sites, except on the dot at the extreme right of the chain which contains $3$ electrons).
 Starting at $t=0$, the electron is emitted, and appears as a left-propagating
wavepacket, which initially has an exponential profile. As it propagates along the chain, this wavepacket broadens due to the
non-linear spectrum of the chain. At $t=100=T/2$ the emission of a hole starts, with similar properties as the electron.

Performing a Fourier transform for each of the one-electron states gives us access to the density in the wave-vector space (or $k$ space)
as a function of time, which is shown on the right part of Fig.~\ref{fig:Density}. Initially, the system in equilibrium, at zero temperature,
has all the states for $|k| < k_{max}$ occupied (occupation 1, the ``Fermi sea''), while all the other states are empty. For $t>0$, the electron
emission shows up as a well-defined peak above the Fermi sea. Similarly, the emission of the hole for $t>100$ appears as a dip in the Fermi sea. Both structures manifest at the same distance from $k_F$, confirming that the electron and hole are emitted with energies symmetric with respect to the Fermi level.

\begin{figure}
\includegraphics[width=4.cm]{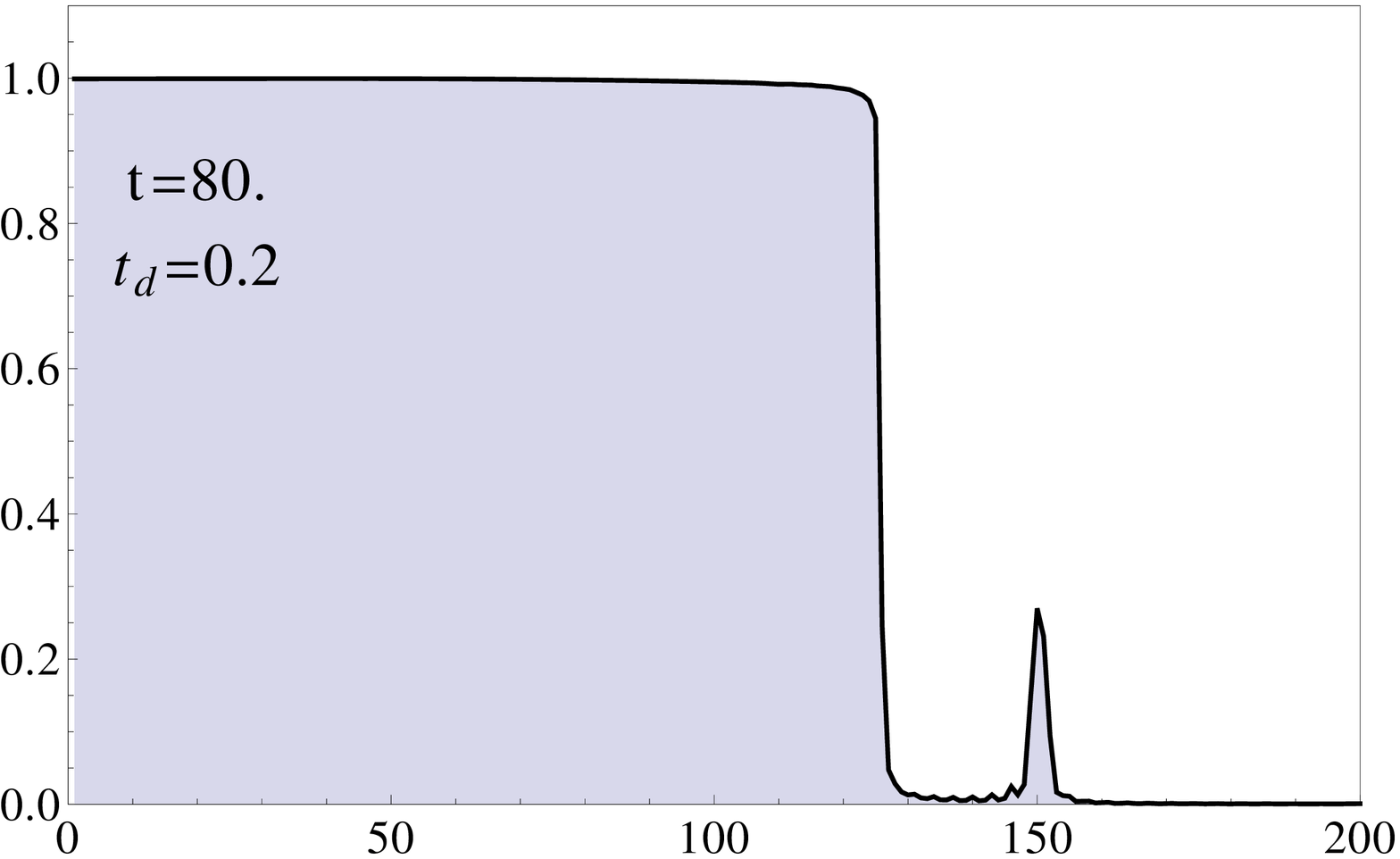} 
\includegraphics[width=4.cm]{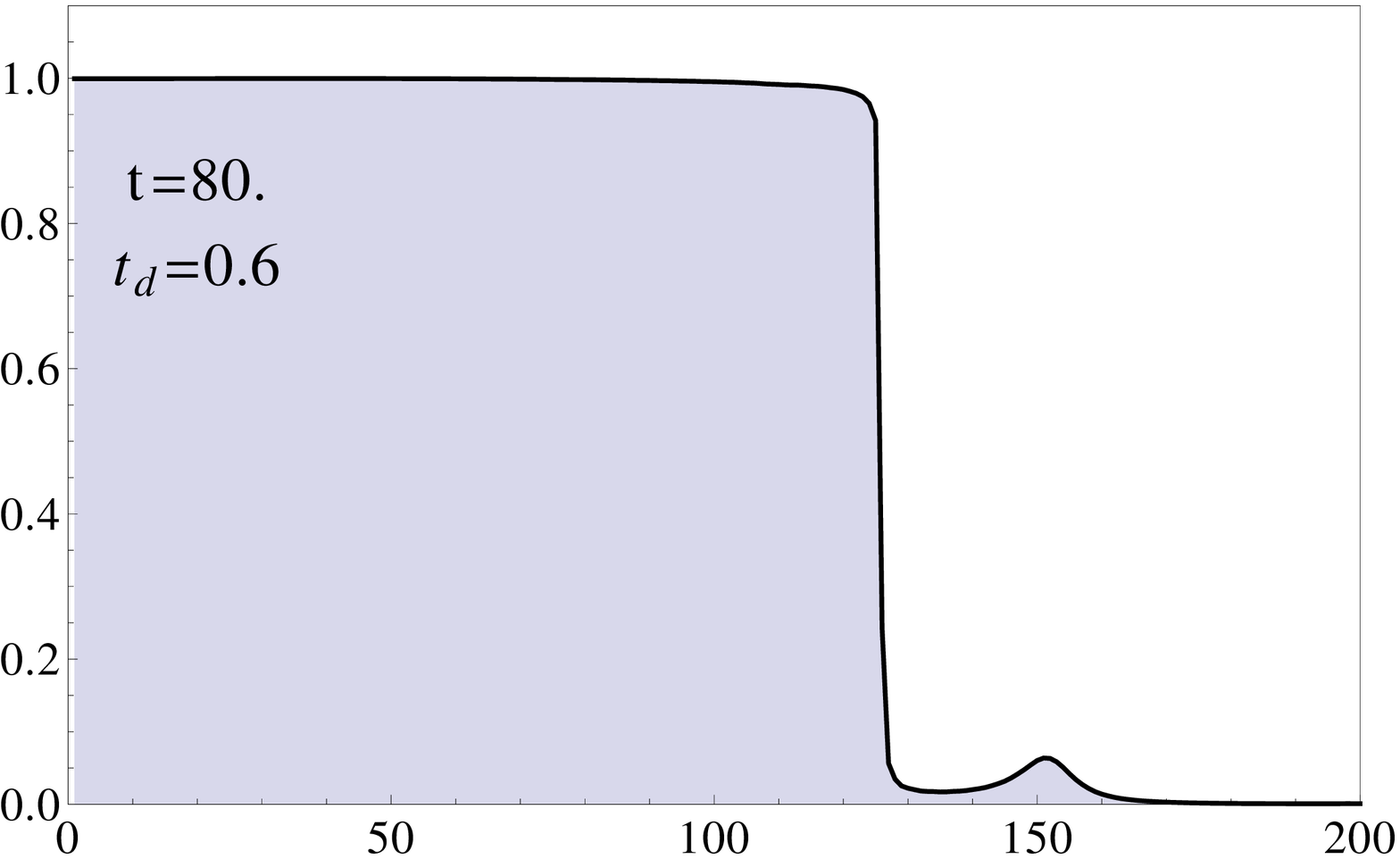} 
\caption{(color online) Effect of the dot tunneling amplitude $t_d$ on the electron peak in momentum space, for $t_d=0.2$ (left) and $t_d=0.6$ (right),
for a chain of 500 sites, with other parameters as in Fig.~\ref{fig:Density}. A smaller transparency leads to a narrower peak well-separated from the occupied states, while a larger transparency gives a broader peak which tends to merge with the occupied states.}
\label{fig:fermipeaks}
\end{figure}

Looking at the density in $k$-space for different tunneling amplitudes $t_d$ gives us information about the properties of the emitted electron, 
and about the ``quality'' of the electron emission. Fig.~\ref{fig:fermipeaks} shows the electron peak for a small tunneling amplitude $t_d=0.2$ 
and a larger one $t_d=0.6$ (and other parameters as in Fig.~\ref{fig:Density}). A small transparency leads to a narrow peak which is well
separated from the Fermi sea, while a larger transparency creates a broader peak
 which tends to merge with the Fermi sea.\cite{grenier_single-electron_2011}

\subsection{Time-dependent current}
\label{sec:avI}

The current coming out of the dot contains similar information to the time-dependent charge on the dot, as it is given by the time derivative of this charge. 
We recall that the current is computed on a link very close to the dot (a few sites away), in order to minimize the effect of the non-linear spectrum of the chain, which tends to spread the wavepacket propagating along the chain during its travel.

\begin{figure}[t]
\centerline{\includegraphics[width=8.cm]{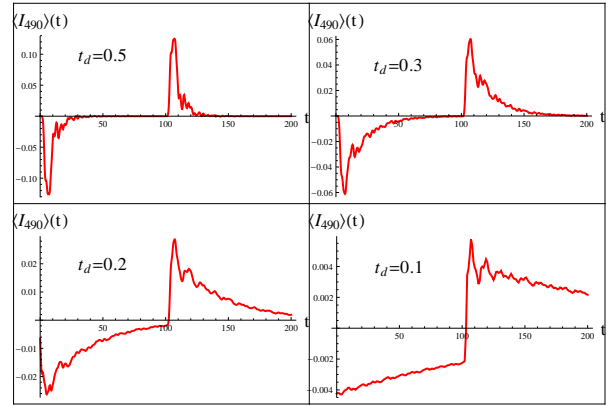}} 
\caption{(color online) Mean current $\langle I \rangle$ between sites 490 and 491
 as a function of time when $V(t)$ is applied for different tunneling to the dot $t_d$
 at zero temperature. Data obtained for a chain of length $L=500$, with a period $T=200$.}
\label{fig:Ioft}
\end{figure}

Fig.\ref{fig:Ioft} shows the current for the same parameters as Fig.\ref{fig:Qdynamic}, computed from Eq.~(\ref{eq:currentT0}) between
sites $490$ and $491$ for a chain of length $500$.
For  $t_d = 0.5 \mbox{ and } 0.3$, the current shows a dip followed by a peak, due to the passage of the electron then the hole.
The peaks have an exponential profile with some oscillations which correspond to the ones present in the charge $\langle Q \rangle (t)$ (we could check explicitly that computing $d\langle Q \rangle (t)/dt$ gives the same peaks). As $t_d$ is reduced, the width of the electron and hole wavepackets increases, and becomes
larger than the half period. This is apparent on the plots for $t_d=0.2$ and $t_d=0.1$, where the current does not reach zero
at $t=100=T/2$. There, when the external drive switches from $V_{+}$ to $V_{-}$, and the hole emission begins, the sign of the current changes abruptly. 
Bear in mind that we had to start the emission at time $t_{in}<0$ (see Sec.~\ref{sec:Qt}) in order to be immediately in the periodic regime at $t=0$. This has the undesired consequence to reduce the oscillations of the current in the first half-period compared to the following ones, as can be seen on the bottom right plot of Fig.~\ref{fig:Ioft} (the current for later periods -not shown- is similar to the one of the second half-period). 
This is because, when using a non zero $t_{in}$, the charge has the correct ``periodic'' value at $t=0$, but there is no abrupt change of the potential at that time unlike what happens at $t=T/2, T,...$. The values for the current during the first half-period are thus somewhat inaccurate. For this reason, in the calculation of the 
current autocorrelations (see below), we use the current starting from the second half-period only.

\subsection{Current autocorrelations}  

 The current correlations at finite frequency contain more information
 about the single emission process than the mean current or the mean charge do.
 We have computed the autocorrelation for currents on the same link:
 \begin{multline}
 S_{k,k} (t_1,t_2)=\left \langle \Psi \middle| I_k(t_1) I_k(t_2) \middle| \Psi \right \rangle 
 \\ -\left \langle \Psi \middle| I_k(t_1) \middle| \Psi \right \rangle
    \left \langle \Psi \middle| I_k(t_2) \middle| \Psi \right \rangle
 \end{multline}
for a link $k$ close to the dot.  Because the applied voltage is periodic in time, $S_{k,k}(t_1,t_2)$ is a periodic function
of the mean time $(t_1+t_2)/2$. By taking the average over one period $T$, we obtain the autocorrelation noise which depends on the time
difference only:
\begin{equation}
S_{k,k}(t) = \overline{S_{k,k}(t',t'+t)}^{t'} .
\label{eq:Stimeaverage}
\end{equation}
Performing a Fourier transform of $S_{k,k}(t)$ gives access to the autocorrelation noise as a function of frequency. This quantity has been
measured experimentally for a frequency of the order of the frequency of the applied voltage $V(t)$, as reported in Ref.~\onlinecite{mahe_current_2010}.
Using a semi-classical model, valid in the optimal emission regime where the Fermi energy of the chain
is halfway between two dot levels, Albert et al.\cite{albert_accuracy_2010} have provided an analytical formula for the
autocorrelation noise as a function of the frequency $\omega$ and the escape time $\tau$ of the electron from the dot:
\begin{equation}
S_{k,k}(\omega) = \frac{2}{T} \tanh\left(\frac{T}{4 \tau}\right) \frac{\omega^2 \tau^2}{1+\omega^2 \tau^2} ,
\label{eq:albert}
\end{equation} 
where $T$ is the period of the applied voltage. A good agreement between this formula and the experimental results was obtained.\cite{mahe_current_2010,albert_accuracy_2010}. 

\begin{figure}
\centerline{\includegraphics[width=7.cm]{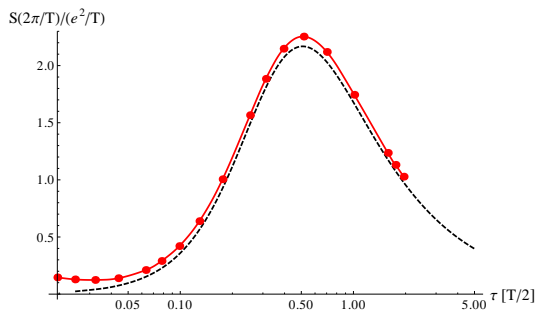}} 
\caption{(color online) Autocorrelation noise $S$ of the current on link 590, taken at the frequency $2 \pi/T$ of the applied potential $V(t)$, as a
function of the escape time from the dot. The noise is expressed in unit of the drive frequency ($1/T$), while the escape
time is in unit of the half-period ($T/2$).
The black dashed curve shows the theoretical prediction of the semi-classical model of Ref.~\onlinecite{albert_accuracy_2010}. The
red points show the results obtained with our calculations, for a chain of length $L=600$ and a period $T=200$.}
\label{fig:SplotL600}
\end{figure}

Fig.~\ref{fig:SplotL600} shows the results obtained for $S_{k,k}(\omega=2\pi/T)$ in units of the driving frequency ($1/T$),
as a function of the escape time of the dot scaled by the half period $\tau/(T/2)$, for a chain of length $L=600$ with a period $T=200$. 
The current correlations were computed at zero temperature, close to the dot, on the link 
between sites $590$ and $591$ (while the dot occupies sites $596$ to $600$).
Note that in this regime, the results are nearly insensitive to temperature, as long as it is much smaller than the level spacing in the dot.
The first observation to be made is that the numerical results 
(red points) are very close to the predicted analytical formula (dashed black curve): the numerical curve is just slightly above the theoretical one (by $\approx 4\%$). 

The curve has a maximum for $\tau \sim T/2$, delimitating two regions which correspond to two different regimes. On the left of the maximum, when $\tau \ll T/2$, the emission time of the electron or hole is much shorter than the half-period, so it is
emitted with probability 1, and the noise comes from the uncertainty in the emission time, which decreases with $\tau$. This is the jitter
regime, or phase noise regime. On the right, when $\tau \gg T/2$, the emission time of the electron or hole is larger that the half-period,
so the emission probability is smaller than 1, which is the cause of the noise. This is thus the shot noise regime, or charge noise regime.
Note that the deviations of our numerical results with respect to the theoretical curve are similar in both regimes. This is a good indication of the reliability of our calculations, in particular that the use of an initial time $t_{in} <0$,
which is only needed in the shot noise regime (see Sec.~\ref{sec:Qt}) gives correct results. 
Larger deviations appear only for very small values of $\tau/(T/2)$, which correspond to the large transparency regime when the escape time from the dot gets close to the time of a round trip inside the dot. The semi-classical model cannot be used in that regime and the analytical formula \eqref{eq:albert} is no longer valid.

We have chosen the period $T=200$ as large as possible for the chain length $L=600$, taking into account the finite size constraints.
Indeed, the maximum time before an emitted wavepacket is reflected at the boundary of the chain and comes back spoiling the measurement of the current, is approximatively $t_{max} \simeq 600$ (because the Fermi velocity is 2 sites per unit of time). The time averaging procedure (Eq.~(\ref{eq:Stimeaverage})) requires at least one period $T$,  while the integration for the Fourier transform needs at least another period $T$.
Moreover, we had to discard the current obtained during the first half-period, as some of its features are different from those of the real periodic current
(see the lower right panel of Fig.~\ref{fig:Ioft}, and discussion at the end of Sec.~\ref{sec:avI}). Taking $T=t_{max}/3\simeq L/3$ is thus close to the maximum value which can reasonably
be taken for the period $T$. Another constraint prevents us from taking arbitrarily large values for the escape time. The magnitude of the initial emission time $t_{in}<0$ that we need to use increases with $\tau$, making it difficult to reach numerically large values of $\tau$, as it imposes to have a larger maximum time $t_{max}$, and thus
a longer chain. This is the reason why the largest value of $\tau/(T/2)$ on Fig.~\ref{fig:SplotL600} is approximatively $2$.

\begin{figure}
\includegraphics[width=4.25cm]{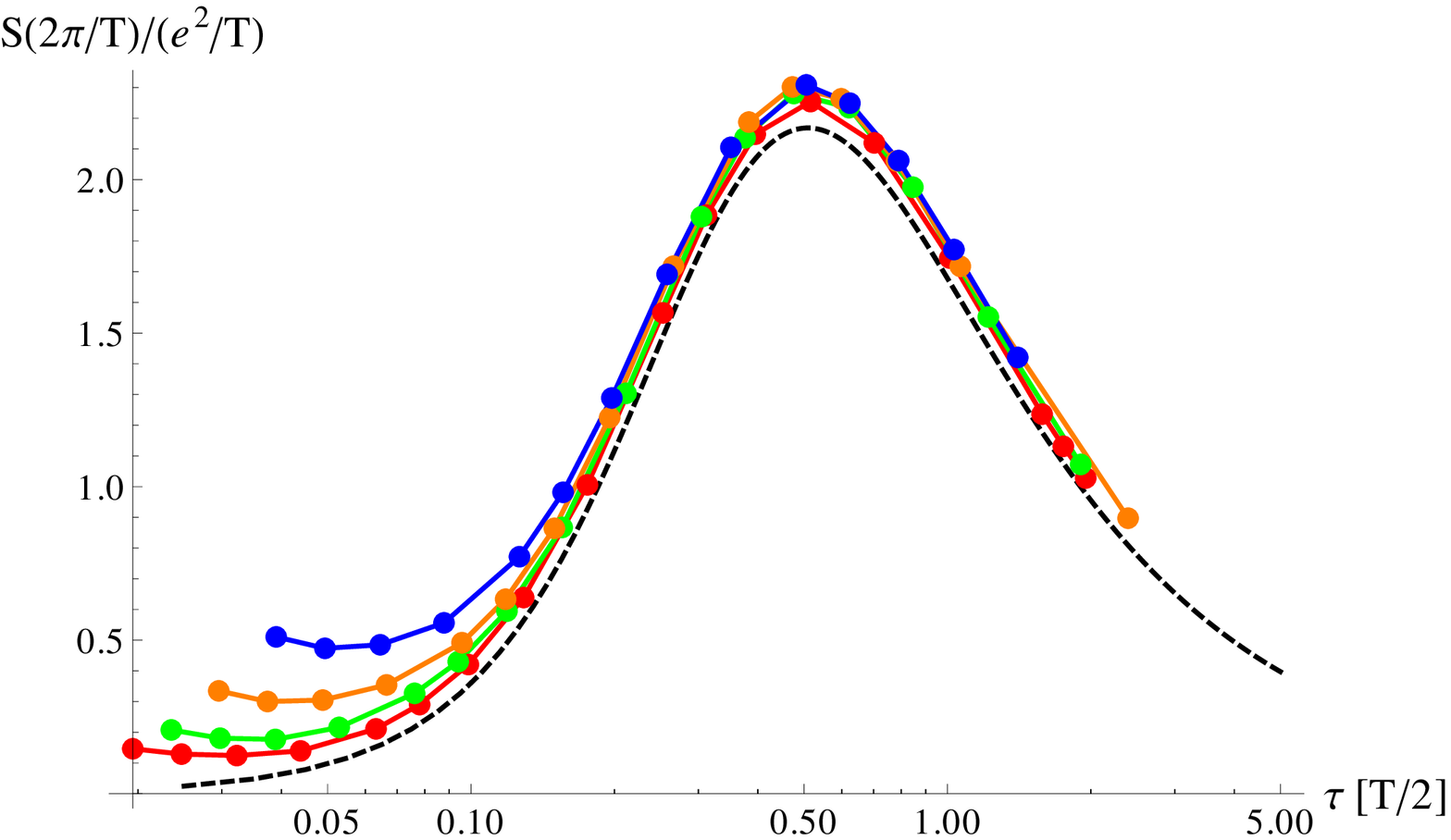} 
\includegraphics[width=4.25cm]{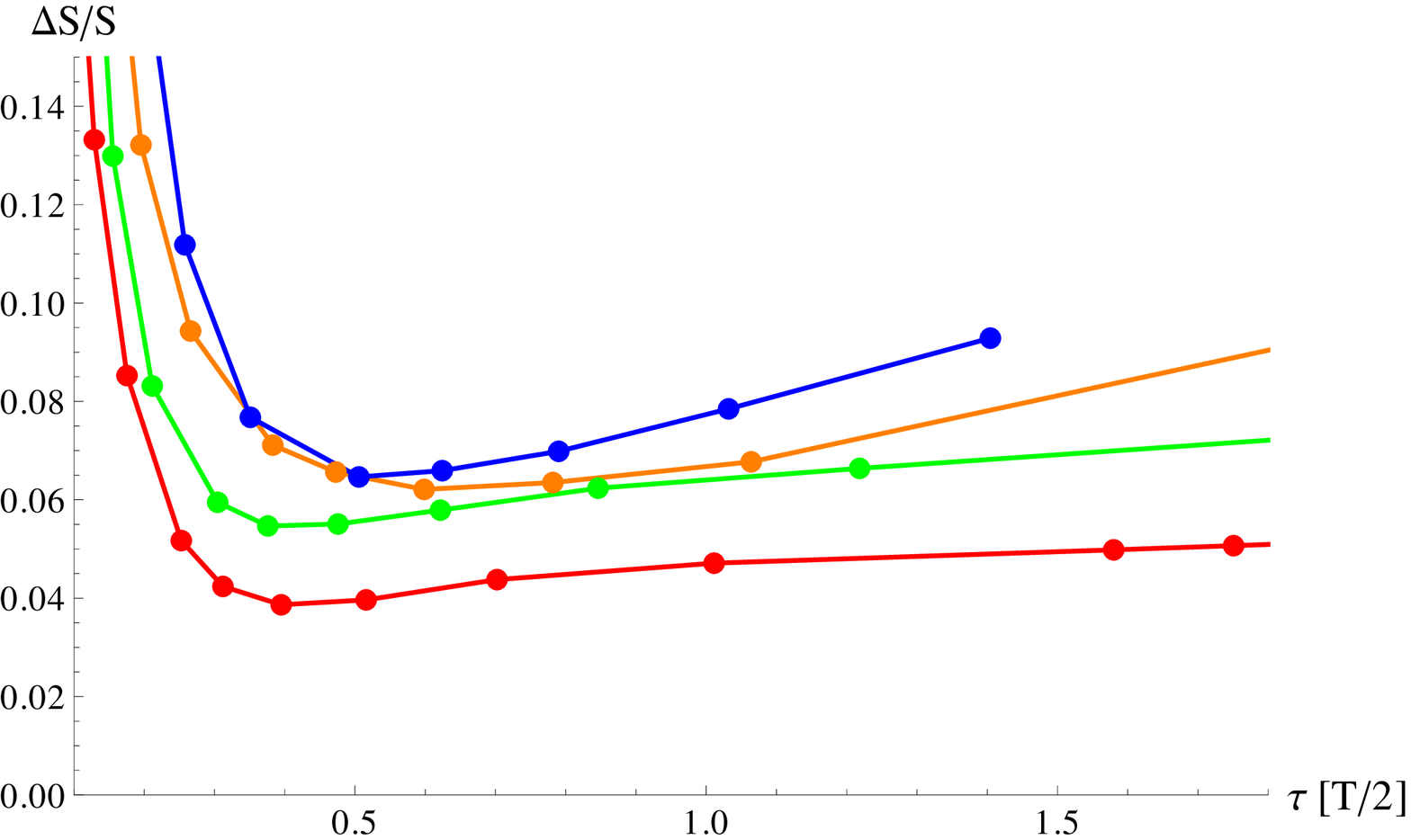} 
\caption{(color online) (Left) Same as Fig.~\ref{fig:SplotL600} for chain lengths $L=300$, $400$, $500$ and $600$ (blue, orange, green and red, or from top to bottom). In each case the period has been chosen as $T=L/3$. Smaller chain gives a higher autocorrelation noise.
(Right) Relative difference $\Delta S / S$ between the computed autocorrelation and the theoretical formula, for the same chain lengths, as a
function of the escape time from the dot. The relative difference decreases when the chain length (and thus the period $T$) increases;
 for the longer chain ($L=600$), the relative difference is about $4 \%$}
 \label{fig:SplotallL}
\end{figure}

In order to investigate the effects of the finite size on the difference between the numerical results and the theoretical curve,
we have computed the current autocorrelation for increasing sizes, $L=300$, $400$ and $500$, taking $T=L/3$ in each case.
As shown on Fig.~\ref{fig:SplotallL} (left), taking a smaller chain, and thus a smaller period $T$, gives a higher current autocorrelation.
 We also see that the deviations from the analytical model for large transparencies (bottom left part of the curves) appears 
 for larger values of $\tau/(T/2)$ for decreasing  chain lengths $L$. This is because
a given value of $\tau/(T/2)$ corresponds to a larger transparency for a smaller chain, which has a smaller $T$.
  
The right plot of Fig.~\ref{fig:SplotallL} shows the relative difference between the computed autocorrelation and the theoretical prediction,
$\Delta S/ S = (S_{(k,k)}(2\pi/T) - S_{theor})/S_{theor}$, as a function of the escape time $\tau$. We see that this relative difference
decreases from the 8-10 \% range for a chain with $L=300$ and $T=100$ to approximatively 4\% for the $L=600$ chain with $T=200$.  One can
thus expect that using a longer chain with a larger $T$ should bring the numerical results closer to the theoretical prediction, 
although a systematic difference, independent of the chain length, cannot be excluded from our results.

\begin{figure}
\centerline{ \includegraphics[width=2.cm]{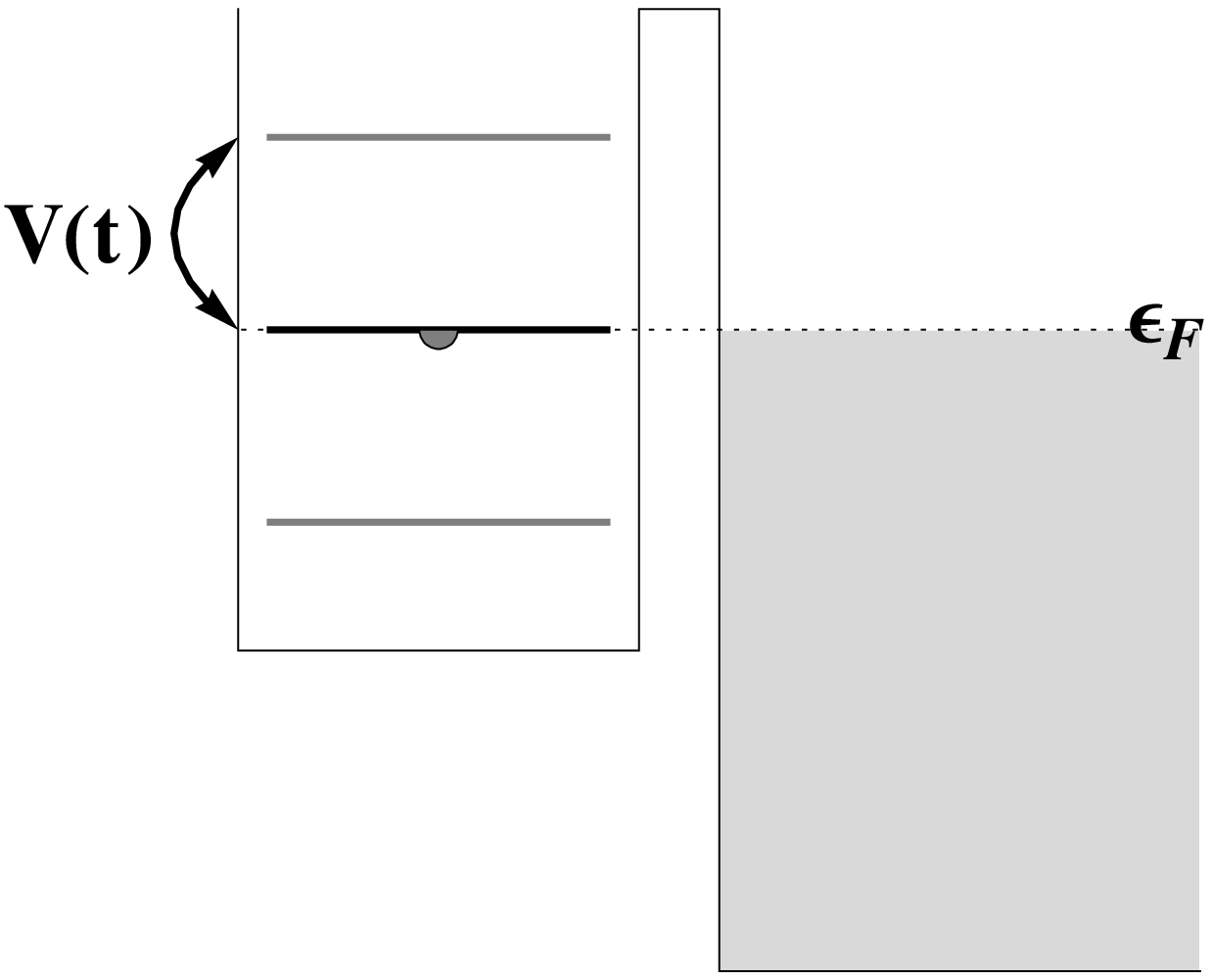} 
  \includegraphics[width=7.cm]{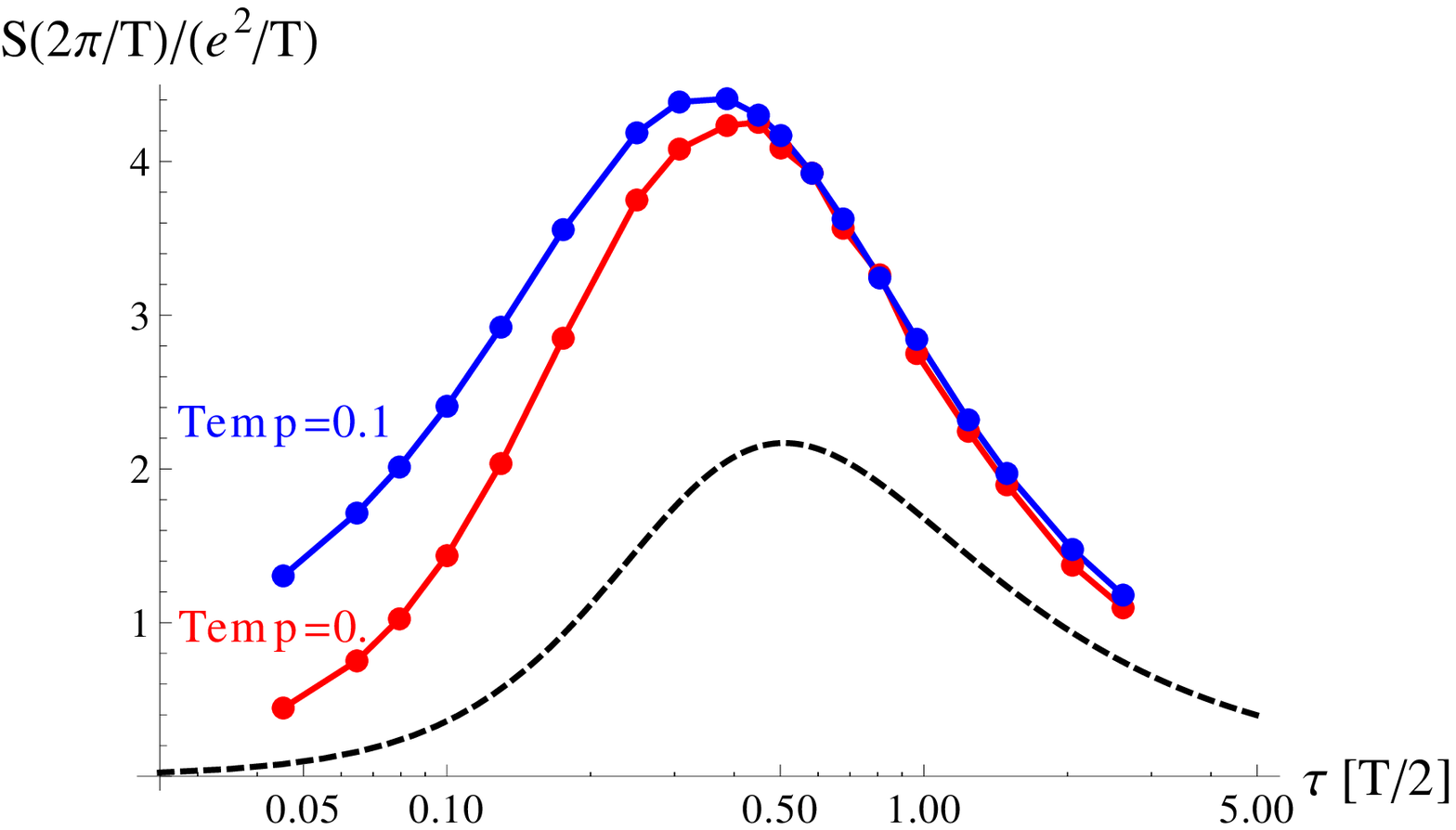}} 
\caption{(color online) Autocorrelation noise in a same setup as in Fig.~\ref{fig:SplotL600}, in the resonant regime: dot levels
are put at the Fermi energy by the external potential (see left panel). The curves with dots show the results obtained at temperature 0 (red) and 0.1 (blue). The black dashed curve is the analytical prediction in the optimal regime. The noise in the resonant regime is much higher than in the optimal regime, showing that the resonant regime is not suitable for single electron emission.}
\label{fig:plotSeps0}
\end{figure}

Our simulations allow us to compute easily the autocorrelation noise for different operating regimes, which cannot be described by
the semi-classical model. As an example, we have considered the resonant regime, where for the two values of $V(t)$ there is a dot level
resonant with the Fermi energy (see left panel of Fig.~\ref{fig:plotSeps0}). This regime is not expected to provide good quality 
emission of single electron, as two levels are contributing equally, at the same time, to the electron emission 
(the level at the Fermi energy, and the level above it).
This is clearly visible in the computed autocorrelation noise (right panel of Fig.~\ref{fig:plotSeps0}), which is much larger than the noise
in the optimal regime. This regime is also much more sensitive to the temperature since a dot level is always at the Fermi energy: for small emission time $\tau$, the noise obtained at a temperature of 0.1 is larger than the one at zero-temperature.

\section{Conclusions}
\label{sec:conclusion}
In this work, we have shown that real-time simulations on a tight-binding chain are a powerful and versatile tool to study the single-electron
emitter, giving the ability to compute accurately elaborate quantities like the finite frequency current-autocorrelation noise.
 Using a non-interacting system, we have shown that the time evolution can be obtained simply
and efficiently from the evolution of the one electron states, and we have derived the formulas for the useful physical quantities 
(charge on the dot, emitted current, current autocorrelations) in terms of the numerical object describing the one electron states evolution.
This allowed us to study the characteristics of the single-electron emitter as a function of different parameters (dot transparency,
period of the AC drive, temperature, ...), giving us access at the same time to illustrative visualizations (current,
time-dependent density along the chain, etc.) and providing information about complex quantities.
Our calculations of finite-frequency current-autocorrelation noise are in good agreement
with theoretical predictions in the regime when these are applicable (and thus also with experimental observations), 
and can be used to explore other regimes. We have detailed the limitations due to the finite size of the system.

This work can be extended in several directions. First, for non-interacting electrons, we have only explored a small part of the parameter space
and of the possible manipulations of the system. We have mainly focused in this work on the optimal regime for single-electron emission as well as the less effective resonant regime, but the method can also be used to study in detail other regimes. We have also limited our study to the case of a perfect piecewise constant time-dependent potential. Our approach can easily be used to study
the effect of the experimental limitations on this shape of potential, or to study different types of potential (for example $V(t) \sim \sin(t)$) 
which would lead to the emission of electron with different energy distribution. It also has the ability to simulate a true ``1 shot'' experiment, where a single pulse (rather than a periodic drive) is applied on the dot to 
emit one electron, starting from an equilibrium situation. 

Our method can also be extended to study more elaborate setups based on the manipulation of the emitted single electron. For example,
a weak link in the middle of the chain, which would only partially transmit an incoming wavepacket, 
would be equivalent to a quantum point contact. This could be used to perform a real-time study of tomography 
protocol proposals,\cite{grenier_single-electron_2011} 
or to consider more complex setups like interferometers.\cite{haack_coherence_2011}
Placing dots at both end of the chain  could also be used to study two-electron interferences 
processes.\cite{olkhovskaya_shot_2008,juergens_single-particle_2011}

Secondly, this work can be seen as a first step towards the study of the single electron emitter taking into account electronic interactions.
As it shows that using a discrete tight-binding model allows to get a faithful and accurate physical description of the single electron emitter,
we can consider extending our calculations to more powerful methods like the time-dependent density-matrix renormalization group 
(td-DMRG)~\cite{white_density_1992, cazalilla_time-dependent_2002,white_real-time_2004,daley_time-dependent_2004,dias_da_silva_transport_2008,
heidrich-meisner_real-time_2009,branschaedel_conductance_2010,ulrich_density-matrix_2011} 
which can include electronic interactions. This work gives the necessary understanding of the role and limitations of the different parameters in the system (chain size, dot size, period of the AC drive, etc). Using td-DMRG (or similar techniques), we could
 study the impact on the emitted electron of strong Coulomb interactions inside the dot. We could also consider electronic density-density 
 interactions between two edge states, which should be experimentally relevant as the experiments have been achieved so far with edge states of the Quantum Hall effect with a filling factor $\nu=2$, where two edge states are propagating.

{\it Note added.} A  preprint (Ref.~\onlinecite{parmentier_current_2011}) has recently appeared, where  
details of the Floquet scattering theory for the same setup are presented, together with the semiclassical model.

\acknowledgments We gratefully acknowledge useful discussions with G. F\`eve, E. Bocquillon, P. Degiovanni and C. Grenier.  This work was supported
by the ANR grant ``1shot'' (ANR-2010-BLANC-0412).
 
\bibliography{mybib1}{}

\begin{thebibliography}{30}
\expandafter\ifx\csname natexlab\endcsname\relax\def\natexlab#1{#1}\fi
\expandafter\ifx\csname bibnamefont\endcsname\relax
  \def\bibnamefont#1{#1}\fi
\expandafter\ifx\csname bibfnamefont\endcsname\relax
  \def\bibfnamefont#1{#1}\fi
\expandafter\ifx\csname citenamefont\endcsname\relax
  \def\citenamefont#1{#1}\fi
\expandafter\ifx\csname url\endcsname\relax
  \def\url#1{\texttt{#1}}\fi
\expandafter\ifx\csname urlprefix\endcsname\relax\def\urlprefix{URL }\fi
\providecommand{\bibinfo}[2]{#2}
\providecommand{\eprint}[2][]{\url{#2}}

\bibitem[{\citenamefont{Sohn et~al.}(1997)\citenamefont{Sohn, Kouwenhoven, and
  Schön}}]{sohn_mesoscopic_1997}
\bibinfo{author}{\bibfnamefont{L.~L.} \bibnamefont{Sohn}},
  \bibinfo{author}{\bibfnamefont{L.~P.} \bibnamefont{Kouwenhoven}},
  \bibnamefont{and} \bibinfo{author}{\bibfnamefont{G.}~\bibnamefont{Schön}},
  \emph{\bibinfo{title}{Mesoscopic Electron Transport {(NATO} Science Series
  E:}} (\bibinfo{publisher}{Springer}, \bibinfo{year}{1997}),
  \bibinfo{edition}{1st} ed., ISBN \bibinfo{isbn}{0792347374}.

\bibitem[{\citenamefont{Bouchiat et~al.}(2005)\citenamefont{Bouchiat, Gefen,
  Gu{\'e}ron, Montambaux, and Dalibard}}]{houches2005}
\bibinfo{editor}{\bibfnamefont{H.}~\bibnamefont{Bouchiat}},
  \bibinfo{editor}{\bibfnamefont{Y.}~\bibnamefont{Gefen}},
  \bibinfo{editor}{\bibfnamefont{S.}~\bibnamefont{Gu{\'e}ron}},
  \bibinfo{editor}{\bibfnamefont{G.}~\bibnamefont{Montambaux}},
  \bibnamefont{and} \bibinfo{editor}{\bibfnamefont{J.}~\bibnamefont{Dalibard}},
  eds., \emph{\bibinfo{title}{Nanophysics: Coherence and Transport, {\'E}cole
  d'{\'e}t{\'e} de Physique des Houches Session LXXXI}}
  (\bibinfo{publisher}{Elsevier}, \bibinfo{year}{2005}).

\bibitem[{\citenamefont{Onac et~al.}(2006)\citenamefont{Onac, Balestro, van
  Beveren, Hartmann, Nazarov, and Kouwenhoven}}]{onac_using_2006}
\bibinfo{author}{\bibfnamefont{E.}~\bibnamefont{Onac}},
  \bibinfo{author}{\bibfnamefont{F.}~\bibnamefont{Balestro}},
  \bibinfo{author}{\bibfnamefont{L.~H.~W.} \bibnamefont{van Beveren}},
  \bibinfo{author}{\bibfnamefont{U.}~\bibnamefont{Hartmann}},
  \bibinfo{author}{\bibfnamefont{Y.~V.} \bibnamefont{Nazarov}},
  \bibnamefont{and} \bibinfo{author}{\bibfnamefont{L.~P.}
  \bibnamefont{Kouwenhoven}}, \bibinfo{journal}{Phys. Rev. Lett.}
  \textbf{\bibinfo{volume}{96}}, \bibinfo{pages}{176601}
  (\bibinfo{year}{2006}).

\bibitem[{\citenamefont{{Zakka-Bajjani}
  et~al.}(2007)\citenamefont{{Zakka-Bajjani}, S{\'e}gala, Portier, Roche,
  Glattli, Cavanna, and Jin}}]{zakka-bajjani_experimental_2007}
\bibinfo{author}{\bibfnamefont{E.}~\bibnamefont{{Zakka-Bajjani}}},
  \bibinfo{author}{\bibfnamefont{J.}~\bibnamefont{S{\'e}gala}},
  \bibinfo{author}{\bibfnamefont{F.}~\bibnamefont{Portier}},
  \bibinfo{author}{\bibfnamefont{P.}~\bibnamefont{Roche}},
  \bibinfo{author}{\bibfnamefont{D.~C.} \bibnamefont{Glattli}},
  \bibinfo{author}{\bibfnamefont{A.}~\bibnamefont{Cavanna}}, \bibnamefont{and}
  \bibinfo{author}{\bibfnamefont{Y.}~\bibnamefont{Jin}},
  \bibinfo{journal}{Phys. Rev. Lett.} \textbf{\bibinfo{volume}{99}},
  \bibinfo{pages}{236803} (\bibinfo{year}{2007}).

\bibitem[{\citenamefont{Aguado and Kouwenhoven}(2000)}]{aguado_double_2000}
\bibinfo{author}{\bibfnamefont{R.}~\bibnamefont{Aguado}} \bibnamefont{and}
  \bibinfo{author}{\bibfnamefont{L.~P.} \bibnamefont{Kouwenhoven}},
  \bibinfo{journal}{Phys. Rev. Lett.} \textbf{\bibinfo{volume}{84}},
  \bibinfo{pages}{1986} (\bibinfo{year}{2000}).

\bibitem[{\citenamefont{Zazunov et~al.}(2007)\citenamefont{Zazunov, Creux,
  Paladino, Cr{\'e}pieux, and Martin}}]{zazunov_detection_2007}
\bibinfo{author}{\bibfnamefont{A.}~\bibnamefont{Zazunov}},
  \bibinfo{author}{\bibfnamefont{M.}~\bibnamefont{Creux}},
  \bibinfo{author}{\bibfnamefont{E.}~\bibnamefont{Paladino}},
  \bibinfo{author}{\bibfnamefont{A.}~\bibnamefont{Cr{\'e}pieux}},
  \bibnamefont{and} \bibinfo{author}{\bibfnamefont{T.}~\bibnamefont{Martin}},
  \bibinfo{journal}{Phys. Rev. Lett.} \textbf{\bibinfo{volume}{99}},
  \bibinfo{pages}{066601} (\bibinfo{year}{2007}).

\bibitem[{\citenamefont{F{\`e}ve et~al.}(2007)\citenamefont{F{\`e}ve, Mah{\'e},
  Berroir, Kontos, Pla{\c c}ais, Glattli, Cavanna, Etienne, and
  Jin}}]{feve_-demand_2007}
\bibinfo{author}{\bibfnamefont{G.}~\bibnamefont{F{\`e}ve}},
  \bibinfo{author}{\bibfnamefont{A.}~\bibnamefont{Mah{\'e}}},
  \bibinfo{author}{\bibfnamefont{J.}~\bibnamefont{Berroir}},
  \bibinfo{author}{\bibfnamefont{T.}~\bibnamefont{Kontos}},
  \bibinfo{author}{\bibfnamefont{B.}~\bibnamefont{Pla{\c c}ais}},
  \bibinfo{author}{\bibfnamefont{D.~C.} \bibnamefont{Glattli}},
  \bibinfo{author}{\bibfnamefont{A.}~\bibnamefont{Cavanna}},
  \bibinfo{author}{\bibfnamefont{B.}~\bibnamefont{Etienne}}, \bibnamefont{and}
  \bibinfo{author}{\bibfnamefont{Y.}~\bibnamefont{Jin}},
  \bibinfo{journal}{Science} \textbf{\bibinfo{volume}{316}},
  \bibinfo{pages}{1169 } (\bibinfo{year}{2007}).

\bibitem[{\citenamefont{Leicht et~al.}(2011)\citenamefont{Leicht, Mirovsky,
  Kaestner, Hohls, Kashcheyevs, Kurganova, Zeitler, Weimann, Pierz, and
  Schumacher}}]{leicht_generation_2011}
\bibinfo{author}{\bibfnamefont{C.}~\bibnamefont{Leicht}},
  \bibinfo{author}{\bibfnamefont{P.}~\bibnamefont{Mirovsky}},
  \bibinfo{author}{\bibfnamefont{B.}~\bibnamefont{Kaestner}},
  \bibinfo{author}{\bibfnamefont{F.}~\bibnamefont{Hohls}},
  \bibinfo{author}{\bibfnamefont{V.}~\bibnamefont{Kashcheyevs}},
  \bibinfo{author}{\bibfnamefont{E.~V.} \bibnamefont{Kurganova}},
  \bibinfo{author}{\bibfnamefont{U.}~\bibnamefont{Zeitler}},
  \bibinfo{author}{\bibfnamefont{T.}~\bibnamefont{Weimann}},
  \bibinfo{author}{\bibfnamefont{K.}~\bibnamefont{Pierz}}, \bibnamefont{and}
  \bibinfo{author}{\bibfnamefont{H.~W.} \bibnamefont{Schumacher}},
  \bibinfo{journal}{Semicond. Sci. Technol.} \textbf{\bibinfo{volume}{26}},
  \bibinfo{pages}{055010} (\bibinfo{year}{2011}).

\bibitem[{\citenamefont{Mah{\'e} et~al.}(2010)\citenamefont{Mah{\'e},
  Parmentier, Bocquillon, Berroir, Glattli, Kontos, Pla{\c c}ais, F{\`e}ve,
  Cavanna, and Jin}}]{mahe_current_2010}
\bibinfo{author}{\bibfnamefont{A.}~\bibnamefont{Mah{\'e}}},
  \bibinfo{author}{\bibfnamefont{F.~D.} \bibnamefont{Parmentier}},
  \bibinfo{author}{\bibfnamefont{E.}~\bibnamefont{Bocquillon}},
  \bibinfo{author}{\bibfnamefont{J.}~\bibnamefont{Berroir}},
  \bibinfo{author}{\bibfnamefont{D.~C.} \bibnamefont{Glattli}},
  \bibinfo{author}{\bibfnamefont{T.}~\bibnamefont{Kontos}},
  \bibinfo{author}{\bibfnamefont{B.}~\bibnamefont{Pla{\c c}ais}},
  \bibinfo{author}{\bibfnamefont{G.}~\bibnamefont{F{\`e}ve}},
  \bibinfo{author}{\bibfnamefont{A.}~\bibnamefont{Cavanna}}, \bibnamefont{and}
  \bibinfo{author}{\bibfnamefont{Y.}~\bibnamefont{Jin}},
  \bibinfo{journal}{Phys. Rev. B} \textbf{\bibinfo{volume}{82}},
  \bibinfo{pages}{201309} (\bibinfo{year}{2010}).

\bibitem[{\citenamefont{Albert et~al.}(2010)\citenamefont{Albert, Flindt, and
  B{\"u}ttiker}}]{albert_accuracy_2010}
\bibinfo{author}{\bibfnamefont{M.}~\bibnamefont{Albert}},
  \bibinfo{author}{\bibfnamefont{C.}~\bibnamefont{Flindt}}, \bibnamefont{and}
  \bibinfo{author}{\bibfnamefont{M.}~\bibnamefont{B{\"u}ttiker}},
  \bibinfo{journal}{Phys. Rev. B} \textbf{\bibinfo{volume}{82}},
  \bibinfo{pages}{041407} (\bibinfo{year}{2010}).

\bibitem[{\citenamefont{Moskalets and
  B\"{u}ttiker}(2002)}]{moskalets_floquet_2002}
\bibinfo{author}{\bibfnamefont{M.}~\bibnamefont{Moskalets}} \bibnamefont{and}
  \bibinfo{author}{\bibfnamefont{M.}~\bibnamefont{B\"{u}ttiker}},
  \bibinfo{journal}{Phys. Rev. B} \textbf{\bibinfo{volume}{66}},
  \bibinfo{pages}{205320} (\bibinfo{year}{2002}).

\bibitem[{\citenamefont{Moskalets and
  B\"{u}ttiker}(2007)}]{moskalets_time-resolved_2007}
\bibinfo{author}{\bibfnamefont{M.}~\bibnamefont{Moskalets}} \bibnamefont{and}
  \bibinfo{author}{\bibfnamefont{M.}~\bibnamefont{B\"{u}ttiker}},
  \bibinfo{journal}{Phys. Rev. B} \textbf{\bibinfo{volume}{75}},
  \bibinfo{pages}{035315} (\bibinfo{year}{2007}).

\bibitem[{\citenamefont{Moskalets et~al.}(2008)\citenamefont{Moskalets,
  Samuelsson, and B\"{u}ttiker}}]{moskalets_quantized_2008}
\bibinfo{author}{\bibfnamefont{M.}~\bibnamefont{Moskalets}},
  \bibinfo{author}{\bibfnamefont{P.}~\bibnamefont{Samuelsson}},
  \bibnamefont{and}
  \bibinfo{author}{\bibfnamefont{M.}~\bibnamefont{B\"{u}ttiker}},
  \bibinfo{journal}{Phys. Rev. Lett.} \textbf{\bibinfo{volume}{100}},
  \bibinfo{pages}{086601} (\bibinfo{year}{2008}).

\bibitem[{\citenamefont{Grenier et~al.}(2011)\citenamefont{Grenier, Herv\'{e},
  Bocquillon, Parmentier, Pla\c{c}ais, Berroir, F\`{e}ve, and
  Degiovanni}}]{grenier_single-electron_2011}
\bibinfo{author}{\bibfnamefont{C.}~\bibnamefont{Grenier}},
  \bibinfo{author}{\bibfnamefont{R.}~\bibnamefont{Herv\'{e}}},
  \bibinfo{author}{\bibfnamefont{E.}~\bibnamefont{Bocquillon}},
  \bibinfo{author}{\bibfnamefont{F.~D.} \bibnamefont{Parmentier}},
  \bibinfo{author}{\bibfnamefont{B.}~\bibnamefont{Pla\c{c}ais}},
  \bibinfo{author}{\bibfnamefont{J.~M.} \bibnamefont{Berroir}},
  \bibinfo{author}{\bibfnamefont{G.}~\bibnamefont{F\`{e}ve}}, \bibnamefont{and}
  \bibinfo{author}{\bibfnamefont{P.}~\bibnamefont{Degiovanni}},
  \bibinfo{journal}{New Journal of Physics} \textbf{\bibinfo{volume}{13}},
  \bibinfo{pages}{093007} (\bibinfo{year}{2011}), ISSN
  \bibinfo{issn}{1367-2630}.

\bibitem[{\citenamefont{Parmentier et~al.}(2011)\citenamefont{Parmentier,
  Bocquillon, Berroir, Glattli, Pla\c{c}ais, F\`{e}ve, Albert, Flindt, and
  B\"{u}ttiker}}]{parmentier_current_2011}
\bibinfo{author}{\bibfnamefont{F.~D.} \bibnamefont{Parmentier}},
  \bibinfo{author}{\bibfnamefont{E.}~\bibnamefont{Bocquillon}},
  \bibinfo{author}{\bibfnamefont{J.~M.} \bibnamefont{Berroir}},
  \bibinfo{author}{\bibfnamefont{D.~C.} \bibnamefont{Glattli}},
  \bibinfo{author}{\bibfnamefont{B.}~\bibnamefont{Pla\c{c}ais}},
  \bibinfo{author}{\bibfnamefont{G.}~\bibnamefont{F\`{e}ve}},
  \bibinfo{author}{\bibfnamefont{M.}~\bibnamefont{Albert}},
  \bibinfo{author}{\bibfnamefont{C.}~\bibnamefont{Flindt}}, \bibnamefont{and}
  \bibinfo{author}{\bibfnamefont{M.}~\bibnamefont{B\"{u}ttiker}},
  \bibinfo{journal}{{arXiv:1111.3136}}  (\bibinfo{year}{2011}).

\bibitem[{\citenamefont{White}(1992)}]{white_density_1992}
\bibinfo{author}{\bibfnamefont{S.~R.} \bibnamefont{White}},
  \bibinfo{journal}{Phys. Rev. Lett.} \textbf{\bibinfo{volume}{69}},
  \bibinfo{pages}{2863} (\bibinfo{year}{1992}).

\bibitem[{\citenamefont{Cazalilla and
  Marston}(2002)}]{cazalilla_time-dependent_2002}
\bibinfo{author}{\bibfnamefont{M.~A.} \bibnamefont{Cazalilla}}
  \bibnamefont{and} \bibinfo{author}{\bibfnamefont{J.~B.}
  \bibnamefont{Marston}}, \bibinfo{journal}{Phys. Rev. Lett.}
  \textbf{\bibinfo{volume}{88}}, \bibinfo{pages}{256403}
  (\bibinfo{year}{2002}).

\bibitem[{\citenamefont{White and Feiguin}(2004)}]{white_real-time_2004}
\bibinfo{author}{\bibfnamefont{S.~R.} \bibnamefont{White}} \bibnamefont{and}
  \bibinfo{author}{\bibfnamefont{A.~E.} \bibnamefont{Feiguin}},
  \bibinfo{journal}{Phys. Rev. Lett.} \textbf{\bibinfo{volume}{93}},
  \bibinfo{pages}{076401} (\bibinfo{year}{2004}).

\bibitem[{\citenamefont{Daley et~al.}(2004)\citenamefont{Daley, Kollath,
  Schollw{\"o}ck, and Vidal}}]{daley_time-dependent_2004}
\bibinfo{author}{\bibfnamefont{A.~J.} \bibnamefont{Daley}},
  \bibinfo{author}{\bibfnamefont{C.}~\bibnamefont{Kollath}},
  \bibinfo{author}{\bibfnamefont{U.}~\bibnamefont{Schollw{\"o}ck}},
  \bibnamefont{and} \bibinfo{author}{\bibfnamefont{G.}~\bibnamefont{Vidal}},
  \bibinfo{journal}{Journal of Statistical Mechanics: Theory and Experiment}
  \textbf{\bibinfo{volume}{2004}}, \bibinfo{pages}{P04005}
  (\bibinfo{year}{2004}), ISSN \bibinfo{issn}{1742-5468}.

\bibitem[{\citenamefont{Dias~da Silva et~al.}(2008)\citenamefont{Dias~da Silva,
  {Heidrich-Meisner}, Feiguin, Bosser, Martins, Anda, and
  Dagotto}}]{dias_da_silva_transport_2008}
\bibinfo{author}{\bibfnamefont{L.~G. G.~V.} \bibnamefont{Dias~da Silva}},
  \bibinfo{author}{\bibfnamefont{F.}~\bibnamefont{{Heidrich-Meisner}}},
  \bibinfo{author}{\bibfnamefont{A.~E.} \bibnamefont{Feiguin}},
  \bibinfo{author}{\bibfnamefont{C.~A.} \bibnamefont{Bosser}},
  \bibinfo{author}{\bibfnamefont{G.~B.} \bibnamefont{Martins}},
  \bibinfo{author}{\bibfnamefont{E.~V.} \bibnamefont{Anda}}, \bibnamefont{and}
  \bibinfo{author}{\bibfnamefont{E.}~\bibnamefont{Dagotto}},
  \bibinfo{journal}{Phys. Rev. B} \textbf{\bibinfo{volume}{78}},
  \bibinfo{pages}{195317} (\bibinfo{year}{2008}).

\bibitem[{\citenamefont{{Heidrich-Meisner}
  et~al.}(2009)\citenamefont{{Heidrich-Meisner}, Feiguin, and
  Dagotto}}]{heidrich-meisner_real-time_2009}
\bibinfo{author}{\bibfnamefont{F.}~\bibnamefont{{Heidrich-Meisner}}},
  \bibinfo{author}{\bibfnamefont{A.~E.} \bibnamefont{Feiguin}},
  \bibnamefont{and} \bibinfo{author}{\bibfnamefont{E.}~\bibnamefont{Dagotto}},
  \bibinfo{journal}{Phys. Rev. B} \textbf{\bibinfo{volume}{79}},
  \bibinfo{pages}{235336} (\bibinfo{year}{2009}).

\bibitem[{\citenamefont{Bransch{\" a}del
  et~al.}(2010{\natexlab{a}})\citenamefont{Bransch{\" a}del, Schneider, and
  Schmitteckert}}]{branschaedel_conductance_2010}
\bibinfo{author}{\bibfnamefont{A.}~\bibnamefont{Bransch{\" a}del}},
  \bibinfo{author}{\bibfnamefont{G.}~\bibnamefont{Schneider}},
  \bibnamefont{and}
  \bibinfo{author}{\bibfnamefont{P.}~\bibnamefont{Schmitteckert}},
  \bibinfo{journal}{1004.4178}  (\bibinfo{year}{2010}{\natexlab{a}}).

\bibitem[{\citenamefont{Schollwoeck}(2011)}]{ulrich_density-matrix_2011}
\bibinfo{author}{\bibfnamefont{U.}~\bibnamefont{Schollwoeck}},
  \bibinfo{journal}{Annals of Physics} \textbf{\bibinfo{volume}{326}},
  \bibinfo{pages}{96} (\bibinfo{year}{2011}), ISSN \bibinfo{issn}{0003-4916}.

\bibitem[{\citenamefont{Bransch{\" a}del
  et~al.}(2010{\natexlab{b}})\citenamefont{Bransch{\" a}del, Boulat, Saleur,
  and Schmitteckert}}]{branschaedel_numerical_2010}
\bibinfo{author}{\bibfnamefont{A.}~\bibnamefont{Bransch{\" a}del}},
  \bibinfo{author}{\bibfnamefont{E.}~\bibnamefont{Boulat}},
  \bibinfo{author}{\bibfnamefont{H.}~\bibnamefont{Saleur}}, \bibnamefont{and}
  \bibinfo{author}{\bibfnamefont{P.}~\bibnamefont{Schmitteckert}},
  \bibinfo{journal}{Phys. Rev. B} \textbf{\bibinfo{volume}{82}},
  \bibinfo{pages}{205414} (\bibinfo{year}{2010}{\natexlab{b}}).

\bibitem[{\citenamefont{Bardeen}(1961)}]{bardeen_tunnelling_1961}
\bibinfo{author}{\bibfnamefont{J.}~\bibnamefont{Bardeen}},
  \bibinfo{journal}{Phys. Rev. Lett.} \textbf{\bibinfo{volume}{6}},
  \bibinfo{pages}{57} (\bibinfo{year}{1961}).

\bibitem[{\citenamefont{Ferrer et~al.}(1988)\citenamefont{Ferrer,
  {Mart\'{i}n-Rodero}, and Flores}}]{ferrer_contact_1988}
\bibinfo{author}{\bibfnamefont{J.}~\bibnamefont{Ferrer}},
  \bibinfo{author}{\bibfnamefont{A.}~\bibnamefont{{Mart\'{i}n-Rodero}}},
  \bibnamefont{and} \bibinfo{author}{\bibfnamefont{F.}~\bibnamefont{Flores}},
  \bibinfo{journal}{Phys. Rev. B} \textbf{\bibinfo{volume}{38}},
  \bibinfo{pages}{10113} (\bibinfo{year}{1988}).

\bibitem[{\citenamefont{Cuevas}(1999)}]{PhDCuevas}
\bibinfo{author}{\bibfnamefont{J.~C.} \bibnamefont{Cuevas}}, Ph.D. thesis,
  \bibinfo{school}{Universidad Autonoma de Madrid} (\bibinfo{year}{1999}),
  \bibinfo{note}{appendix. B}.

\bibitem[{\citenamefont{Haack et~al.}(2011)\citenamefont{Haack, Moskalets,
  Splettstoesser, and B\"{u}ttiker}}]{haack_coherence_2011}
\bibinfo{author}{\bibfnamefont{G.}~\bibnamefont{Haack}},
  \bibinfo{author}{\bibfnamefont{M.}~\bibnamefont{Moskalets}},
  \bibinfo{author}{\bibfnamefont{J.}~\bibnamefont{Splettstoesser}},
  \bibnamefont{and}
  \bibinfo{author}{\bibfnamefont{M.}~\bibnamefont{B\"{u}ttiker}},
  \bibinfo{journal}{Phys. Rev. B} \textbf{\bibinfo{volume}{84}},
  \bibinfo{pages}{081303} (\bibinfo{year}{2011}).

\bibitem[{\citenamefont{Ol{\textquoteright}khovskaya
  et~al.}(2008)\citenamefont{Ol{\textquoteright}khovskaya, Splettstoesser,
  Moskalets, and B\"{u}ttiker}}]{olkhovskaya_shot_2008}
\bibinfo{author}{\bibfnamefont{S.}~\bibnamefont{Ol{\textquoteright}khovskaya}},
  \bibinfo{author}{\bibfnamefont{J.}~\bibnamefont{Splettstoesser}},
  \bibinfo{author}{\bibfnamefont{M.}~\bibnamefont{Moskalets}},
  \bibnamefont{and}
  \bibinfo{author}{\bibfnamefont{M.}~\bibnamefont{B\"{u}ttiker}},
  \bibinfo{journal}{Phys. Rev. Lett.} \textbf{\bibinfo{volume}{101}},
  \bibinfo{pages}{166802} (\bibinfo{year}{2008}).

\bibitem[{\citenamefont{Juergens et~al.}(2011)\citenamefont{Juergens,
  Splettstoesser, and Moskalets}}]{juergens_single-particle_2011}
\bibinfo{author}{\bibfnamefont{S.}~\bibnamefont{Juergens}},
  \bibinfo{author}{\bibfnamefont{J.}~\bibnamefont{Splettstoesser}},
  \bibnamefont{and}
  \bibinfo{author}{\bibfnamefont{M.}~\bibnamefont{Moskalets}},
  \bibinfo{journal}{{EPL} {(Europhysics} Letters)}
  \textbf{\bibinfo{volume}{96}}, \bibinfo{pages}{37011} (\bibinfo{year}{2011}),
  ISSN \bibinfo{issn}{0295-5075, 1286-4854}.

\end{thebibliography}

\end{document}